\documentclass[aps,prb,reprint,twocolumn,notitlepage,floatfix,tightenlines,superscriptaddress,longbibliography]{revtex4-1}
\usepackage{amssymb}
\usepackage{amsfonts}
\usepackage{bbm}
\usepackage{amsmath}
\usepackage{graphicx}
\usepackage[caption=false]{subfig}
\usepackage[colorlinks=true,linkcolor=blue,anchorcolor=red,citecolor=blue,urlcolor=blue]{hyperref}
\usepackage{enumitem}

\newcommand\GammaVar{\reflectbox{\rotatebox[origin=c]{180}{$\mathbb L$}}}

\renewcommand{\vec}[1]{\mathbf{#1}}

\newcommand{\veck}{\vec{k}}
\newcommand{\vecq}{\vec{q}}

\newcommand{\Tr}{\mathrm{Tr}}

\newcommand{\LW}{\mathrm{LW}}

\newcommand{\bsigma}{\boldsymbol{\sigma}}

\newcommand{\He}{$^{3}$He}
\newcommand{\RPA}{\mathrm{RPA}}

\newcommand{\heliumthree}{$^3$He}

\makeatletter

\numberwithin{equation}{section}

\usepackage[colorlinks=true,citecolor=blue,linkcolor=blue,urlcolor=blue]{hyperref}

\makeatother

\begin{document}
   
\title{Non-Quantum-Critical Routes to Magnetic Superconductivity}

\author{Zhiqiang Wang}
\email[]{wzhiqustc@ustc.edu.cn}
\affiliation{Hefei National Research Center for Physical Sciences at the Microscale and School of Physical Sciences, University of Science and Technology of China,  Hefei, Anhui 230026, China}
\affiliation{Shanghai Research Center for Quantum Science and CAS Center for Excellence in Quantum Information and Quantum Physics, University of  Science and Technology of China, Shanghai 201315, China}
\affiliation{Hefei National Laboratory, University of  Science and Technology of China, Hefei 230088, China}
\affiliation{Department of Physics and James Franck Institute, University of Chicago, Chicago, IL 60637, USA}
\author{Ke Wang}
\affiliation{Department of Physics and James Franck Institute, University of Chicago, Chicago, IL 60637, USA}
\author{K. Levin}
\email[]{levin@jfi.uchicago.edu}
\affiliation{Department of Physics and James Franck Institute, University of Chicago, Chicago, IL 60637, USA}

\date{\today}

\begin{abstract}
Electronic superconductivity is most commonly understood to be associated with magnetic quantum criticality. This framework is natural when a quantum critical point is present below the peak of the $T_c$
dome, but it is less satisfactory in the many systems 
where electronic superconductivity appears 
without a visible quantum critical point (QCP). Why and where does superconductivity emerge in such cases, given that its condensation energy is generally much smaller than the energy scale of the magnetic order itself?
Here we develop an energetic perspective of this non-quantum-critical route to 
electronic superconductivity. When magnetic order becomes incipient but cannot be fully realized, the 
opening of a pairing gap lowers the free-energy cost of the nearby fluctuating magnetic state. This 
means that pairing
can become
strongest where  
long range order first becomes fragile. As a result, the strongest pairing tendency can 
sometimes occur at the edge of the superconducting dome closest to the loss of magnetism, even when $T_c$ 
 itself is relatively small there.
The relevant organizing principle is therefore not quantum criticality itself, but proximity to unrealized or disappearing magnetic order. Because many unconventional superconductors show no clear QCP, this perspective provides an
alternative framework for understanding the phase diagram of a large class
of electronic superconductors and may help identify new superconducting materials.
\end{abstract}

\maketitle

\section{Introduction}

Magnetic superconductors such as heavy fermions, iron pnictides, cuprates and
others~\cite{Weng2016} are generally associated with a magnetically ordered partner. 
It is understood that this particle-hole partner in some way
is responsible for the attractive interaction leading to superconductivity.
That has been the focus of the bulk of theoretical research in the field
which frames the proximity between superconductivity and
magnetism in terms of magnetic fluctuations acting as a pairing glue 
~\cite{Scalapino1986,Bickers1989a,Monthoux1994,Monthoux2007,Scalapino2012,Varma2012},
especially near a magnetic quantum critical point (QCP). 
But that picture does not by itself explain precisely where or when
superconductivity emerges in the phase diagram and why it so often appears at the point where magnetism weakens and disappears,
in systems without a clear quantum critical point.

One might think that once there is an attractive interaction superconducting order
must set in.
Upon deeper examination this raises a puzzle, however, as the characteristic energy associated
with magnetic order is so much larger than typical condensation energies. One
may wonder why
and where should superconductivity ever have the incentive to emerge when
its energy scales are so small?
The answer proposed here is that, 
provided a pairing glue is available,
superconductivity becomes energetically competitive only
 as magnetic order becomes weaker
and more fragile.
When nearby order becomes attenuated or cannot fully develop, the associated fluctuating state  carries an excess free-energy cost. The opening of a pairing gap
can lower that cost and thereby provide an additional stabilization of the paired phase. In this sense, magnetism does more than furnish a pairing interaction~\cite{Scalapino1986,Bickers1989a,Monthoux1994,Monthoux2007,Scalapino2012,Varma2012}; it creates an energetic drive toward pairing itself. 
This energetic perspective which we develop here has received less
attention in the literature than the source of the pairing attraction. The pairing attraction and the 
energetic drive
to superconductivity are complementary: neither by itself is sufficient to fully explain
experiment.

\begin{figure*}
\centering
\includegraphics[width=4.5in]
{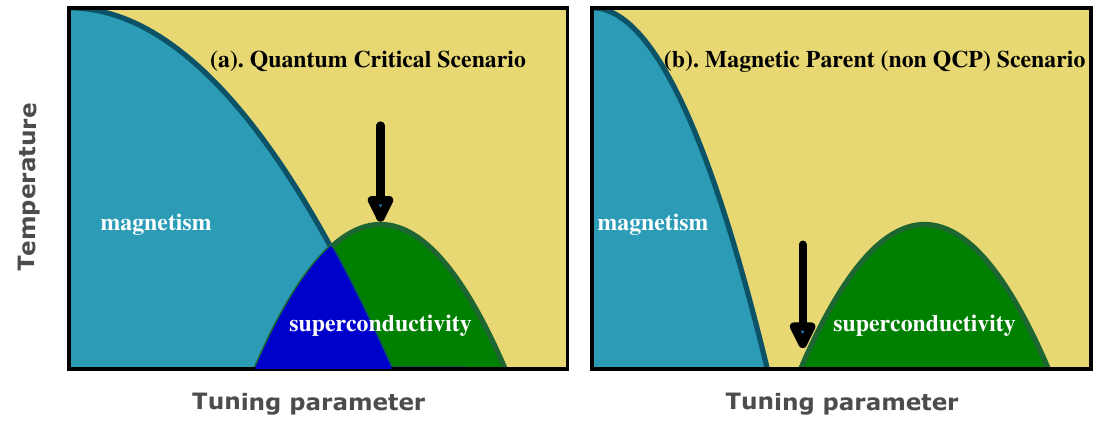}
\caption{Two schematic phase-diagram scenarios for the location of strongest pairing near disappearing 
(long range ordered) magnetism. 
In each panel, the arrow indicates where the pairing is strongest. Panel (a) represents the conventional quantum-critical-point scenario, in which the maximum $T_c$ coincides with the disappearance of magnetic order. Panel (b) represents the non QCP
scenario, more relevant to systems such as the cuprates, where the strongest pairing instead occurs at the edge of the superconducting dome closest to the loss of magnetism in the parent phase. It is there that the largest pairing gap $\Delta$ appears even though $T_c$ is reduced. Although this may seem counterintuitive~\cite{Nozieres1985,Chen2024}, the lower $T_c$ reflects the reduced mobility of strongly bound pairs, as discussed in the text.
}
\label{fig:Fig1b}
\end{figure*}

The more ``standard" approach to magnetic superconductors, however,
invokes a quantum critical point
~\cite{Abanov2003,Chubukov2020,Varma2012,Shibauchi2014,Abrahams2011,Gegenwart2008,Landaeta2018,Landaeta2017}
beneath the superconducting dome, generally near the point of maximal $T_c$, as schematically shown in Fig.~\ref{fig:Fig1b}(a). In that picture, superconductivity is enhanced by critical magnetic fluctuations associated with the loss of long-range order.
Indeed, there is a widespread belief in the community that electronic superconductivity
is generically described by and is equivalent to quantum criticality.
The present paper takes an alternative viewpoint which we argue requires representation
in the literature as there are
many cases in which superconductivity remains tied to disappearing magnetism without a 
clear quantum critical point. This is illustrated by the contrasting situation of Fig.~\ref{fig:Fig1b}(b).
Understanding this prototypical ``non-QCP" phase diagram
is, thus, a central focus of this paper.

Indeed, the cuprate phase diagram
which is similar to 
Fig.~\ref{fig:Fig1b}(b)
serves to illustrate these complexities. There one finds
the maximum of the superconducting dome is well separated from the magnetic parent phase.
Nevertheless, the strongest pairing, 
inferred for example from the size of the measured~
\cite{Damascelli2003} pairing gap $\Delta$, is in the region closest to the
endpoint of the magnetic phase. 
The key question is therefore not whether magnetic fluctuations can mediate BCS pairing---that is already well established---but why, in this class of phase diagrams, the strongest pairing does \emph{not} occur where $T_c$ is largest. We argue that understanding this structure requires two ingredients taken together: 
\begin{itemize}
\item a distinction between pairing strength and superconducting coherence, so that stronger pairing need not imply larger $T_c$; and 
\item a microscopically derived
energetic mechanism showing there is an incentive to pair when magnetism disappears.
\end{itemize}

Earlier work emphasized
~\cite{Nozieres1985}
that in the strong pairing limit there is an important
distinction between pairing strength and coherent order.
$T_c$ is
suppressed as the underlying pairing scale grows. 
This reflects the
impaired mobility~\cite{Nozieres1985,Chen2024} of the Cooper pairs on a lattice and is
a necessary consequence when their binding is strong. It
provides a natural explanation for why the regime of strongest pairing can lie away from the point of maximal transition temperature.

Belonging to the non-QCP scenarios,
there is also an important class of magnetically-driven superconductors
in which no magnetic phase is directly visible in the phase diagram. Examples include superfluid helium-3, where no magnetic order is realized in the liquid state, as well as heavy-fermion materials such as CeCoIn$_5$ and PuCoGa$_5$ 
~\cite{Sidorov2002,Ramshaw2015}.
In such systems, superconductivity is associated not with an observed magnetic phase boundary, but is driven by magnetism that remains incipient or virtual.
These cases sharpen the distinction between the existence of a pairing ``glue" and the separate question of the underlying thermodynamics which favors the paired state.

Thus, this non-QCP structure,
in Fig.~\ref{fig:Fig1b}(b)
is not confined to the cuprates. It is also visible in organic superconductors with very short coherence lengths near the Mott boundary~\cite{Suzuki2022}, and in moiré
superconductors~\cite{Cao2018a,Cao2018b} where the shortest coherence lengths occur 
adjacent to correlated insulators.
While the existence of an attractive interaction 
is necessary, it is not sufficient to explain these phase diagrams.
Playing a key role, as we shall show, is
that the proximity to unrealized order leads to
an extra thermodynamic incentive toward pair formation. 
From the point of view of the phase diagram, the pairing gap, $\Delta$, is
maximal near the attenuating order.

To characterize this incentive more microscopically,
the formalism we use is an RPA-based Luttinger--Ward description
~\cite{Luttinger1960} of the free energy;
this demonstrates that
the energy gain which favors pairing appears as a constructive feedback effect.
The concept of ``feedback" in this paper corresponds to an incentive to pair when
magnetism is attenuated. 
This enters because the pairing gap modifies the same magnetic fluctuations that mediate pairing, and this change lowers their contribution to the free energy. 
In this way, the paired state is ``self-stabilized" by feeding back
on the pairing
fluctuations which initially produce it.
Important here is the necessity
to explain a recurring and longstanding phase-diagram structure shared by several modern classes of unconventional superconductors.

To support this energetics principle 
one must show that the negative sign of this feedback effect is not specific to a particular microscopic model. Earlier work identified feedback mechanisms designed to investigate different
physics
~\cite{Monthoux1994,Scalapino2012,Anderson1973,Brinkman1974,Amin2020,Kozii2019}.
Here we establish within an RPA-Luttinger-Ward approximation
that the associated enhancement of the energy gain arises rather generally for electronically mediated pairing. 
While the detailed form of the susceptibilities and pairing kernels depends on microscopic structure, the basic energetic mechanism is not
tied to the microscopic details of a single model: pairing can be additionally stabilized by lowering the free energy of a nearby but unrealized particle-hole instability.

Thus, rather than focusing on the nature of a magnetic pairing glue, our concern here is the integration
of strong pairing physics into the non-quantum-critical phase diagrams. In many materials the largest gap, shortest
coherence length or highest pairing scale lies closest to the disappearance of magnetic order,
even when the superconducting transition is maximum elsewhere. 
The resulting relationship between regimes of strongest pairing, optimal $T_c$, and the disappearance of magnetism 
leads to 
the unresolved structure of the phase diagram that motivates the present work.

We should note that there are competing ideas which suggest that in many materials a QCP is effectively hidden
because the onset of superconductivity obscures this criticality. 
This possibility has been rather extensively contemplated~\cite{Broun2008} 
and it is claimed that 
at low temperatures~\cite{Sachdev2011} ``bare quantum criticality is invariably preempted 
by another phase",
the most prominent example of which is superconductivity.
To integrate this perspective within the present paper, one needs to address the question of why
superconductivity (with its small condensation energy) should be able to preempt a robust 
magnetic phase. 
Indeed the emphasis here on the energetics is different as it
suggests that superconductivity can only emerge when magnetic order is
weak or close to attenuated.

Throughout this paper we use ``magnetic fluctuations'' in a broad sense, to include strong particle-hole-channel fluctuations associated with nearby symmetry-breaking instabilities, whether spin, charge, or orbital in character. Our focus is on the non-ordered side of such instabilities, not necessarily close to a quantum critical point. The issues we address are therefore broader than the standard quantum-critical scenario, while remaining centered on the recurring empirical proximity between superconductivity and disappearing order.

\subsection{Organization of the paper}

We start in Sect.~\ref{sec:LG} with a brief sketch of Landau-Ginzberg theory which includes both
magnetic and superconducting order
and in this context present intuition about constructive feedback effects in these non-QCP
systems.
This serves to help build a central message of the paper that electronically mediated superconductivity is stabilized
not simply by magnetic fluctuations acting as a pairing attraction, but also by the additional
condensation-energy gain obtained from feeding back on the incipient magnetic ofder. Here we focus on
an ordered magnetization, but later in the paper consider the fluctuating case within a more complete
formalism.

In Sec.~\ref{sec:TcHe3} we present a benchmark example~\cite{Leggett1975}, superfluid helium-3, 
in which there appears to be no QCP and whose pairing mechanism is unusually well constrained by experimentally determined Landau parameters. Helium-3 is particularly instructive because it provides one of the clearest cases in which RPA-like magnetic fluctuations mediate BCS pairing without any accompanying magnetic phase in the liquid state. It therefore serves as a clean prototype for superconductivity or superfluidity near incipient magnetism. 
For readers interested in the more quantitative aspects of this discussion, Appendix~\ref{app:He3_landau} presents additional details on how the Landau parameters arise in paramagnon theory and how they are connected to the superfluid transition temperature $T_c$.

In Sect.~\ref{sec:feedbackLW} we present an investigation of the energetics
of pairing in the presence of a fluctuating magnetism. This leads to
a main formal result and central equation
[Equation~\eqref{eq:PhiRPA4}]. 
Using an RPA-like energy functional within a Luttinger--Ward framework,
it can be seen that feedback from the pairing gap produces a negative correction to the quartic term in the free energy,
corresponding to an additional energy gain. In this way proximity to unrealized magnetic order stabilizes superconductivity.
Appendices~\ref{app:eliashberg_feedback} and \ref{app:mcmillan} show how the same feedback extends naturally to a generalized Eliashberg framework applicable to a wide class of electronically mediated pairing mechanisms.

Finally, in Sect.~\ref{sec:strongpairing} we turn to three specific materials without a clear quantum critical point inside the superconducting dome and show how to understand the phase diagram
which is of the general form of 
Fig.~\ref{fig:Fig1b}(b)
in more concrete terms.
We address two related questions: how such domes arise, and where on the dome the strongest pairing is expected. 

Section~\ref{sec:conclusion} contains our conclusions.

\subsection{Ginzburg-Landau illustration of the energetic mechanism}
\label{sec:LG}
Before turning to the microscopic theory, it is useful to
illustrate the energetics at the level of the simplest
phenomenology. Here we consider Ginzburg-Landau (GL) theory.
The point here is to show how superconductivity gains an additional condensation-energy
contribution by feed back effects due to magnetic order. 
We focus on the ordered state here. In later sections we address some of
these same issues from the perspective of a \textit{fluctuating} ordered state, which is the main focus of
this paper.

Let $M$ denote the magnetic order
parameter and $\Delta$ the superconducting order parameter.
The minimal coupled Landau free energy is
\begin{equation}
F(M,\Delta)
=
a_mM^2
+
\frac{b_m}{2}M^4
+
a_s|\Delta|^2
+
\frac{b_s}{2}|\Delta|^4
+
\gamma M^2|\Delta|^2 ,
\end{equation}
where
\begin{equation}
b_m>0,
\qquad
b_s>0,
\qquad
\gamma>0.
\end{equation}

The positive coupling $\gamma$ describes the competition
between magnetic and superconducting order. Eliminating
the magnetic order parameter by minimizing the free energy,
$M^2
=
-
\frac{a_m+\gamma|\Delta|^2}{b_m}$,
gives an effective superconducting free energy
\begin{equation}
F_{\rm eff}
=
-
\frac{a_m^2}{2b_m}
+
a_s^{\rm eff}|\Delta|^2
+
\frac{b_s^{\rm eff}}{2}
|\Delta|^4,
\end{equation}
where
\begin{equation}
a_s^{\rm eff}
=
a_s
-
\frac{\gamma a_m}{b_m},
\end{equation}
and
\begin{equation}
b_s^{\rm eff}
=
b_s
-
\frac{\gamma^2}{b_m}.
\end{equation}

The negative correction to $b_s$
 is the phenomenological manifestation of the constructive feedback effect.
Two conclusions immediately follow.
First, static magnetic order suppresses superconductivity
through the quadratic coupling. Since $a_m<0$ inside the
magnetic phase,
\begin{equation}
a_s^{\rm eff}
=
a_s
+
\frac{\gamma|a_m|}{b_m}
>
a_s.
\end{equation}
As magnetic order weakens,
$a_m\rightarrow0^-$,
so that this quadratic suppression disappears.

Second, the self consistent adjustment of the magnetic order after superconductivity
develops lowers the quartic coefficient,
\begin{equation}
\delta b
=
-
\frac{\gamma^2}{b_m}
<
0.
\end{equation}

Consequently, the superconducting condensation energy,
\begin{equation}
E_{\rm cond}
=
-
\frac{\left(a_s^{\rm eff}\right)^2}
{2b_s^{\rm eff}},
\end{equation}
is increased because
$b_s^{\rm eff}
<
b_s.$

Thus the magnetic sector plays two distinct roles.
Static magnetic order suppresses superconductivity through
the quadratic coupling, while its self-consistent adjustment after pairing
develops lowers the free energy and provides additional stabilization of the
superconducting state. Although the GL analysis above applies only to the
magnetically ordered phase, we show below that the same stabilization mechanism
also operates in the paramagnetic phase. For this purpose, we use a more
microscopic formalism that naturally incorporates Cooper pairing mediated by
magnetic fluctuations.

\section{Helium-3: A Prototype Non-Quantum-Critical Magnetic Superconductor}
\label{sec:TcHe3}

Before developing a microscopic theory of the feedback mechanism, it is useful
to consider the canonical example of magnetically mediated pairing without either
a quantum critical point or magnetic order in the liquid phase: superfluid
\heliumthree. In this prototype system, pairing is mediated by spin fluctuations,
and $T_c$ is highest near the solidification boundary even though the liquid phase
contains no magnetic QCP. Helium-3 also provides a natural setting for introducing
the paramagnon vertex functions used later in the Luttinger--Ward formulation.

We stress that \heliumthree\ offers an important advantage over most solid-state systems.
Because its normal state is well described by Landau Fermi-liquid theory, the pairing interaction and transition temperature can be related semi-quantitatively to experimentally determined interaction parameters. This allows one to examine, in a relatively controlled microscopic setting, how magnetic-fluctuation-induced pairing arises without requiring either a magnetic phase in the paired fluid or a conventional magnetic quantum critical point. 
For this reason, \heliumthree\ serves here not merely as a historical example, but as a benchmark system.

\begin{figure*}
\begin{center}
\includegraphics[width=0.98\linewidth,clip]
{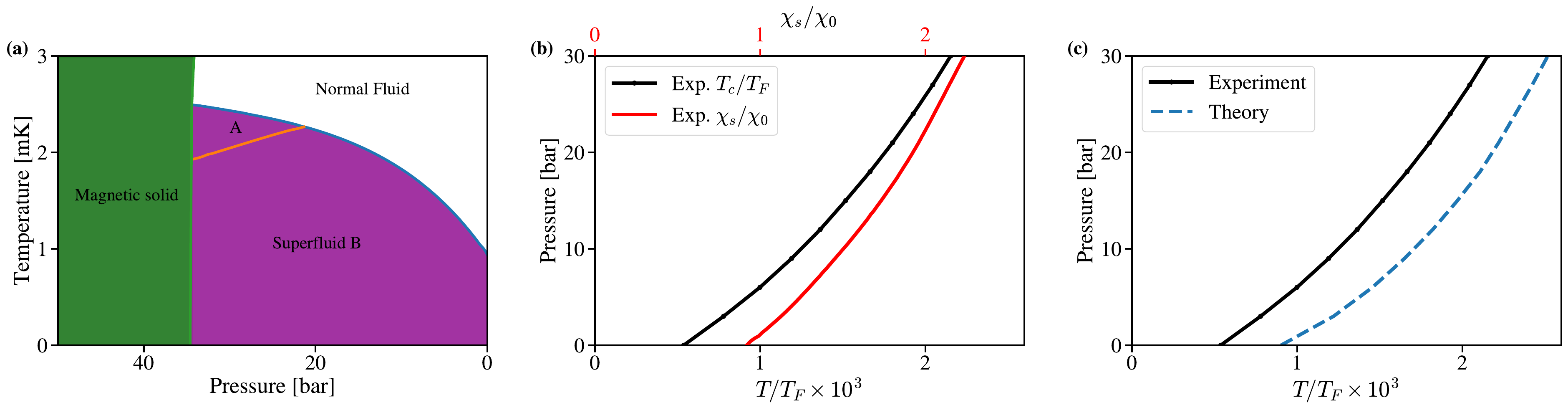}
\caption{Helium-3 phase diagrams. (a) Temperature versus pressure phase diagram for helium-3. Here ``A" refers to the so-called superfluid A phase.
Note that the pressure axis is plotted slightly unconventionally as pressure decreases from the left to right. 
It should be noted that for helium-3 there is a first order
transition from the liquid to the magnetically ordered solid~\cite{captionnote}.
(b) Comparison of the experimental spin
susceptibility $\chi_s$ as a function of pressure (plotted in the vertical axis) with the superfluid transition temperature $T_c$.
(c) Comparison between the theoretical superfluid transition temperature $T_c$ (blue dashed line) predicted by paramagnon theory,
as given by
~\cite{Levin1979a,Levin1983}
$T_c/T_F = 0.1 e^{ - 9/ A_1^s}$,
with the experimental~\cite{Dobbs2000} $T_c$ (black line with dots).
Here, $T_F$ is the Fermi temperature and $A_1^s$ is a pressure dependent Landau Fermi liquid parameter (see text).
At pressures higher than those
plotted in (b) and (c), the system enters the solid phase, where superfluidity ends and magnetism is observed below
a temperature~\cite{Lee1997} of around $1$mK
(for the pressure range shown in the phase diagram in panel (a)).
}
\label{fig:Fig2}
\end{center}
\end{figure*}

The role of magnetism in superfluid \heliumthree\ is to some extent intertwined with the onset of 
BCS-type superfluidity. This is most apparent at higher pressures where the liquid approaches solidification\cite{Cross1985}.
The superfluid transition temperature reaches its maximum, forming the apex of a dome-like structure [Fig.~\ref{fig:Fig2}(a)], precisely as the system is pushed toward the magnetically ordered solid. Equally significant is the evidence for fluctuating or incipient ferromagnetism within the liquid itself, seen in the strong enhancement of the normal-state Pauli spin susceptibility, specific heat, and neutron-scattering response\cite{Levin1983,Leggett1975}. As shown in Fig.~\ref{fig:Fig2}(b), the Pauli susceptibility rises with pressure in close parallel with the transition temperature. Taken together, these observations show that the proximity of superfluidity to magnetism in \heliumthree\ is not incidental: the superfluid phase is strongest precisely where magnetic correlations become most pronounced, even though no magnetic phase appears in the liquid.

Stimulated by observations such as those in Fig.~\ref{fig:Fig2}(a) and Fig.~\ref{fig:Fig2}(b), Anderson and Brinkman proposed that the pairing interaction responsible for superfluidity in \heliumthree\ arises from the exchange of ferromagnetic spin fluctuations\cite{Anderson1973,Brinkman1974}. In this paramagnon picture, the same fluctuations that signal incipient ferromagnetism provide the attractive interaction leading to $p$-wave, spin-triplet pairing.
Helium-3 is therefore important for the present discussion not only because it offers a classic example of electronically mediated pairing, but also because it shows that strong pairing near disappearing magnetism does not require either a magnetic phase boundary within the paired fluid itself or a conventional quantum critical point.

Fig.~\ref{fig:Fig2}(a) shows that the superfluid transition temperature is largest near the pressure at which magnetism appears in the solid. The liquid-to-solid transition is first order, and the relevant magnetic order occurs only after the system leaves the liquid altogether~\cite{Lee1997}. A closely analogous phase-diagram structure appears in certain organic superconductors, such as $\kappa$-(BEDT-TTF)$_2X$~\cite{Nam2007}. For this reason, \heliumthree\ is useful not only because it shares the microscopic pairing mechanism of correlated electronic superconductors, but because it provides an especially transparent benchmark for the broader phase-diagram logic emphasized here: pairing can be strongest in
proximity to a nearby but unrealized magnetic state, even when the phase diagram itself contains neither a resolved magnetic order nor a magnetic quantum critical point.

These observations motivate a more quantitative analysis of pairing in \heliumthree. 
Owing to the exceptional accuracy of its Landau-theory description in the normal state, the effective pairing interaction can be related directly to experimentally extracted Landau parameters, and the resulting transition temperature can then be compared semi-quantitatively with experiment~\cite{Levin1979a,Levin1983}.

\subsection{Transition temperature in \texorpdfstring{$^3$}{3}He via magnetic fluctuations and RPA theory}
\label{sec:TcHe3sub}

The prevailing microscopic picture of superfluidity in \heliumthree\ is that it is 
driven by RPA-like ferromagnetic spin fluctuations. Although alternative viewpoints exist, most notably Vollhardt's ``almost localized'' scenario~\cite{Vollhardt1984} and the Bedell--Pines polarization-potential approach~\cite{Bedell1980}, the paramagnon framework is particularly useful here because it makes explicit how the enhanced magnetic response of the normal liquid is tied to the emergence of spin-triplet, $p$-wave pairing. Our aim is not to review this literature in detail, but rather to isolate the ingredients most relevant to the central theme of this paper: the same collective fluctuations that promote pairing also renormalize the normal state, and must themselves be modified once the paired state forms.

A special advantage of \heliumthree\ is that its relevant interaction parameters are known much more quantitatively than in most electronic systems. In particular, the pressure-dependent Landau scattering amplitudes $A_\ell^{s,a}$ are reasonably well constrained experimentally. As a result, both the pairing interaction and the transition temperature can be estimated with relatively little phenomenological freedom. Liquid \heliumthree\ therefore provides a rare semi-quantitative example of fluctuation-mediated pairing that can be connected rather directly to measurable Fermi-liquid parameters.

As a simplified description of the short-range part of the van der Waals interaction in liquid \heliumthree, we begin with a Hubbard-like contact repulsion,
\begin{align}
\hat H
=
I \sum_{\veck,\veck'} \sum_{\vec q}
c^\dagger_{\veck+\vec q,\uparrow} c_{\veck,\uparrow}
c^\dagger_{\veck'-\vec q,\downarrow} c_{\veck',\downarrow},
\label{eq:Hubbard}
\end{align}
where $I>0$ is the interaction strength. Within the paramagnon approach, this repulsive interaction generates an effective dynamical interaction $\hat H^{\rm eff}$ that depends on the transferred four-momentum $q=(\Omega,\mathbf q)$. A spin-rotation-invariant summation of bubble and ladder diagrams within the random phase approximation (RPA) yields
\begin{align}
\hat H^{\rm eff}
&=
\frac12 \sum_{k,k',q}
\sum_{\{\alpha,\beta,\gamma,\delta\} = \{\uparrow, \downarrow \}}
\bigl[
-J(q)\,\boldsymbol{\sigma}_{\alpha\beta}\!\cdot\!\boldsymbol{\sigma}_{\gamma\delta}  \nonumber \\
& +V(q)\,\delta_{\alpha\beta}\delta_{\gamma\delta}
\bigr]
c^\dagger_{k+q,\alpha}
c^\dagger_{k'-q,\gamma}
c_{k',\delta}
c_{k,\beta},
\label{eq:Heff}
\end{align}
with
\begin{subequations}
\begin{align}
J(q) &= \frac12 \frac{I}{1-I\chi_0(q)},\\
V(q) &= \frac12 \frac{I}{1+I\chi_0(q)},
\end{align}
\end{subequations}
where $\chi_0(q)$ is the Lindhard function of the noninteracting normal-state Fermi liquid. Here, $-J(q)$ denotes the spin-exchange part of the interaction, while $V(q)$ is the spin-independent part. 
In this section, as well as in Appendix~\ref{app:He3_landau}, it is convenient to regard the frequency as real; elsewhere in the paper, the frequency component of a four-momentum is understood to be Matsubara unless stated otherwise.

To connect $\hat H^{\rm eff}$ to experiment, we evaluate the low-energy quasiparticle vertex in the static forward-scattering limit of the particle-hole channel, defined~\cite{Abrikosov1963}
by $|\mathbf q|\to 0$ and $|\Omega|/(v_F|\mathbf q|)\to 0$. Projection onto the spin-symmetric ($s$) and spin-antisymmetric ($a$) states, followed by decomposition into angular momentum channel $\ell$, then yields the Landau quasiparticle scattering amplitudes $A_\ell^{s,a}$. In particular,
\begin{align}
A_1^s = \frac{N_F}{Z}\left(9J_1-3V_1\right),
\label{eq:As}
\end{align}
where $N_F$ is the density of states per spin at the Fermi energy, $J_\ell$ and $V_\ell$ are the Legendre components of the static interactions $J(\Omega=0,|\mathbf q|)$ and $V(\Omega=0,|\mathbf q|)$, and $Z=1+\lambda_Z$ is the quasiparticle frequency-renormalization factor, identified with the effective-mass enhancement $m^*/m$ within the momentum-independent self-energy approximation used here. Appendix~\ref{app:He3_landau} gives the derivation of Eq.~\eqref{eq:As} and its relation to the interaction $\hat{H}^{\text{eff}}$.

The same effective interaction can also be projected into the spin-triplet $p$-wave Cooper channel to estimate the superfluid transition temperature. 
Assuming that the pairing gap scale is small compared with the Fermi energy, the dimensionless pairing strength parameter may be written as
(for details see Appendix~\ref{app:He3_landau})
\begin{equation}
\lambda_\Delta^{\ell=1}=N_F\left(J_1-V_1\right).
\label{eq:lambdaDelta}
\end{equation}
Following McMillan \cite{McMillan1968} the resulting transition temperature is then
\begin{equation}
T_c \propto e^{-Z/\lambda_\Delta^{\ell=1}}
      = e^{-(1+\lambda_Z )/\lambda_\Delta^{\ell=1}},
\label{eq:2.3}
\end{equation}
where $\lambda_Z=Z-1$ denotes the mass renormalization of Landau quasiparticles. 
Numerical Eliashberg calculations\cite{Levin1978} further indicate a characteristic energy cutoff of order $0.1\,T_F$ (Anderson and Brinkman\cite{Brinkman1974} estimated a somewhat smaller value, $\sim 0.05\,T_F$), giving
\begin{equation}
T_c/T_F \approx 0.1\, e^{-Z/\lambda_\Delta^{\ell=1}}.
\label{eq:Tc2}
\end{equation}

When magnetic fluctuations dominate, so that $V_1$ can be neglected compared with $J_1$, 
Eqs.~\eqref{eq:As}, \eqref{eq:lambdaDelta}, and \eqref{eq:Tc2} combine to give the simple estimate\cite{Levin1979a}
\begin{equation}
T_c/T_F \approx 0.1\, e^{-9/A_1^s}.
\label{eq:Tc}
\end{equation}
This result is useful for two reasons. First, it relates the pairing scale directly to an experimentally constrained Landau parameter. 
Second, it makes explicit that the same RPA-based interaction responsible for the enhanced magnetic response of the normal liquid also provides the attraction in the triplet pairing channel. 
Using more recent pressure-dependent data for the Landau parameters\cite{Dobbs2000}, especially $A_1^s$, 
Eq.~\eqref{eq:Tc} yields a semi-quantitative estimate of $T_c$ in reasonable agreement with experiment, as shown in Fig.~\ref{fig:Fig2}(c).

It is notable that the mass renormalization factor 
$Z$ enters $T_c$ on the same footing as the pairing attraction itself as is evident in Eq.~\eqref{eq:2.3}, where the same fluctuation spectrum contributes both to the pairing kernel $\lambda_\Delta^{\ell=1}$ and to the mass renormalization $\lambda_Z$. The latter is not a minor correction. Anderson and Brinkman\cite{Anderson1973} were among the first to estimate $T_c$ for $p$-wave superfluidity in \heliumthree\ within a spin-fluctuation picture, but in 
the language of McMillan they neglected $\lambda_Z$. Later work\cite{Levin1979a} showed that this omission is substantial: $\lambda_Z$ is very large, even exceeding $\lambda_\Delta^{\ell=1}$. Including it reduces the effective coupling from the bare value $\lambda_\Delta^{\ell=1}\sim 4.5$ to the much smaller ratio $\lambda_\Delta^{\ell=1}/(1+\lambda_Z)\sim 0.2$, bringing the estimated transition temperature into much better agreement with experiment.

There is, of course, an important formal caveat. The Eliashberg framework used to estimate $T_c$ relies in its original form on Migdal's theorem, which is not generally valid for electronically mediated pairing.\cite{Hertz1976} Vertex corrections, negligible in the phonon problem, should in principle be included. Even so, Eliashberg-based approaches remain useful here.
We now expand our understanding beyond \heliumthree\ and examine how the development of 
pairing feeds back on the non-superconducting fluctuation sector and thereby 
provides an incentive towards pair formation.

\section{Constructive Feedback (Pairing Incentives) in Electronically Driven Superconductors}
\label{sec:feedbackLW}

\subsection{Luttinger--Ward formalism}
\label{sec:LW}

We now confront a central puzzle to address more generally why 
the strongest pairing scale or pairing gap follows the disappearance of nearby order. In some instances
the onset of strongest pairing need not be related to the maximum
$T_c$. In \heliumthree~\ which has a BCS-like character, we see that these are connected
and that the phase diagram in 
Fig.~\ref{fig:Fig2}(a)
shows the magnetism disappearing 
where the pairing and critical temperature are largest.

We recast the problem in a free-energy language.
To see how this energetic perspective is complementary to, but distinct from, the determination of 
$T_c$, note that within a Landau--Ginzburg formulation the transition temperature is determined by the quadratic term in the superconducting order parameter, or equivalently by the linearized gap equation, whereas the stability of the ordered state and its condensation energy depend on quartic and higher-order terms in the free-energy expansion. Our purpose in this section is to make that higher-order effect explicit: whenever superconductivity feeds back on the fluctuation spectrum that contributes to pairing, an additional condensation-energy gain results.

A convenient starting point is the Luttinger--Ward (LW) functional~\cite{Luttinger1960} for the grand-canonical thermodynamic potential $\Omega_{\rm LW}$. Although originally formulated for an interacting Fermi liquid in the normal state, the LW construction can be generalized to superconductors~\cite{Rainer1976,Haussmann2007,Kita2011}. In general,
\begin{align}
\Omega_{\rm LW}
&=
-\frac{1}{2}\Tr\!\left[
\ln\!\left(-\hat G_0^{-1}+\hat\Sigma\right)
+\hat\Sigma \hat G
\right]
+\Phi
\label{eq:LW1} \\
&\equiv \Omega_{\rm el}+\Phi ,
\label{eq:LW2}
\end{align}
where, for brevity, we suppress the arguments $(i\omega_n,\mathbf{k})$ of the Green's functions and self-energies. The symbol $\Tr$ denotes a functional trace, including the Matsubara sum, momentum integration, and the trace over both spin and Nambu particle--hole space.

In Eq.~\eqref{eq:LW1}, $\hat G_0(i\omega_n,\mathbf{k})$ is the noninteracting Green's function, $\hat\Sigma(i\omega_n,\mathbf{k})$ the self-energy, and $\hat G(i\omega_n,\mathbf{k})$ the full Green's function. Each is a matrix in spin and Nambu space. For the moment, we restrict attention to a single-band setting, so that spin is the only internal degree of freedom besides particle--hole structure. 
The functional 
$\Phi$ 
formally comprises the sum of all closed, linked skeleton diagrams~\cite{Luttinger1960}. In practice, however, this exact sum is essentially impossible to evaluate, so one must work within an appropriate approximation.

The power of the Luttinger--Ward approach is that it is a ``$\Phi$-derivable" theory. When $\Phi$ is approximated by a specified set of skeleton diagrams and the self-energy is obtained self-consistently by functional differentiation, the construction respects the associated conservation laws and thermodynamic identities~\cite{Baym1962,Kadanoff1962}.
Incorporating the RPA within this framework makes it possible to treat the pairing self-energy and the magnetic fluctuations as interlocked quantities.
As a result, feedback effects emerge naturally from the free-energy structure, rather than being introduced as external corrections. 
Although the RPA treatment is approximate, the energetic tendency it captures can be more general: weakening a nearby correlated magnetic state, even an insulating one, can enhance superconductivity.

As a concrete representative of the broader class of particle--hole fluctuations~\cite{Brinkman1968,Doniach1966,Nakajima1973,Brinkman1974}, we first take $\Phi$ to be the ferromagnetic paramagnon functional.
Within this approximation,
\begin{align}
\Phi_{\RPA} = \frac{1}{2}\,\Tr\!\left[\ln\!\left(1-I\hat{\chi}\right)\right].
\label{eq:PhiRPA}
\end{align}
Here $I$ is the Hubbard contact interaction strength introduced in Eq.~\eqref{eq:Hubbard}, $\hat{\chi}\equiv \hat{\chi}(i\Omega_m,\vecq)$ is the spin-susceptibility tensor, and $\Tr$ now denotes the trace over tensor indices together with the integration over bosonic frequency--momentum space. In this form, $\Phi_{\RPA}$ is the paramagnon free energy discussed in Ref.~\onlinecite{Brinkman1974} for superfluid \He. It is the magnetic analog of the phonon free-energy contribution in Eliashberg theory,
$
\Phi_{\mathrm{phonon}}=\frac{1}{2}\Tr[\ln D^{-1}(i\Omega_m,\vecq)],
$
with $D(i\Omega_m,\vecq)$ the phonon propagator.

\begin{figure}[h]
\includegraphics[width=3.5in,clip]{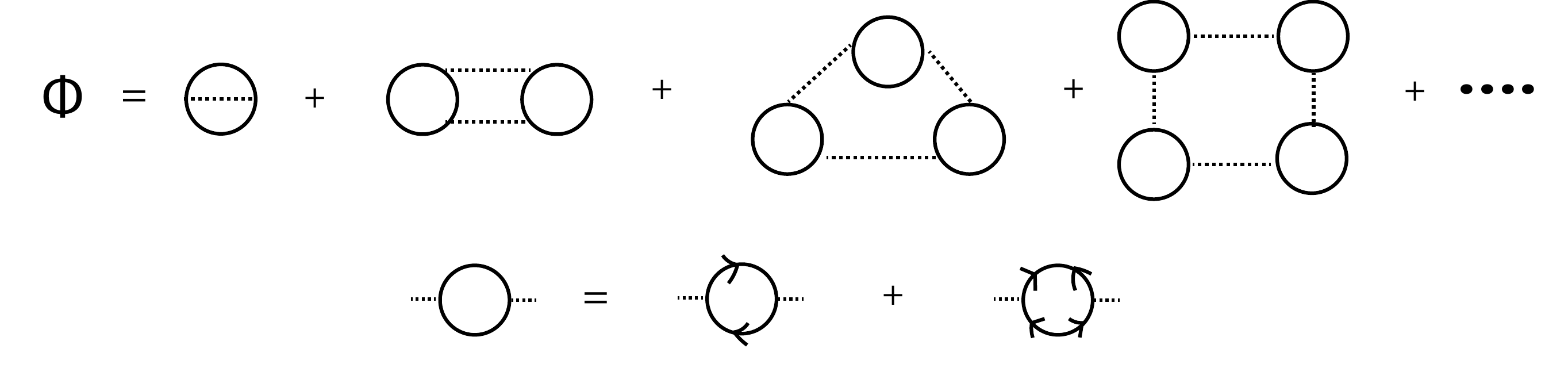}
\caption{Schematic diagrammatic contributions to the Luttinger--Ward functional $\Phi$ within a self-consistent RPA approximation.
The solid line represents the full Green's function, which depends on self-energies.
The dashed line represents a bare spin-exchange interaction, proportional to $-I$ for the Hubbard model in Eq.~\eqref{eq:Hubbard}. Here $I$ is the Hubbard interaction strength.
While a general expansion of $\Phi$ for a Hubbard model also contains diagrammatic contributions
from density--density as well as particle--particle scattering channels~\cite{Bickers1989},
these are neglected in our RPA approximation. This approximation is justified when the system is close to a magnetic
or spin-density-wave instability, so that this diagrammatic sum dominates $\Phi$.
More precisely, $\Phi$ results from the sum of ring and particle--hole exchange diagrams; see Refs.~\onlinecite{Brinkman1968,Doniach1966,Nakajima1973}. 
}
\label{fig:LW}
\end{figure}

Within the paramagnon model, $\Phi$ includes ring and particle--hole exchange diagrams summed to all orders, while all other diagrams are neglected
~\footnote{
In Refs.~\onlinecite{Brinkman1974,Doniach1966} the magnetic-fluctuation contribution is written as
\(\Phi_{\RPA}=\tfrac{1}{2}\Tr[\ln(1-I\hat{\chi})+I\hat{\chi}]\),
so that the fluctuation contribution begins at $O(I^2)$. The added term $I\hat{\chi}$ subtracts the first-order Hartree--Fock contribution from the fluctuation functional and avoids double counting. Here we use Eq.~\eqref{eq:PhiRPA} without that subtraction because the feedback term of interest is determined by the second functional derivative with respect to \(\hat{\chi}\), which is unaffected by a term linear in \(\hat{\chi}\); in addition, near magnetic ordering the dominant physics is controlled by the logarithmic collective term.
},
as shown schematically in Fig.~\ref{fig:LW}. If one focuses on a particular particle--hole channel, such as spin exchange or density fluctuations, this amounts to a self-consistent random phase approximation. For our purposes, the main virtue of this approximation is that it treats the electronic and fluctuation contributions within a common free-energy functional while retaining the collective enhancement associated with proximity to magnetic order, or more generally to a nearby particle--hole instability.

As shown in Appendix~\ref{app:eliashberg_feedback}, just as in phonon-mediated superconductivity, exchange of the paramagnons described by $\Phi_{\RPA}$ generates an attractive interaction between electrons, with matrix kernel
\begin{equation}
\hat{\Gamma}(q)=-\frac{I}{2}\big[1-I\hat{\chi}(q)\big]^{-1}.
\end{equation}
The crucial point is that both $\hat{\chi}$ and $\hat{\Gamma}$, and hence the pairing interaction itself, depend on the superconducting order that they help produce. For a unitary spin-triplet pairing state~\cite{Brinkman1974}, 
\begin{align}
\chi_{ij}(q)
&   = \int_k   \bigg[ -  G(k_-) G(k_+)  + \sum_{\gamma=1}^3 F_\gamma(k_-) F_\gamma^*(k_+) \bigg] \delta_{ij}  \nonumber \\
&  -  F_i(k_-) F_j^*(k_+) - F_j(k_-) F_i^*(k_+).
\label{eq:chisc}
\end{align}
Here we use the four-vector notation
$k=(i\omega_n,\veck)$,
$q=(i\Omega_m,\vecq)$, and
$k_{\pm}=(i\omega_n \pm i\Omega_m/2,\veck \pm \vecq/2)$,
together with the shorthand
$\int_k = T \sum_{\omega_n} \int d\veck/(2\pi)^3$.
In Eq.~\eqref{eq:chisc}, $G$ is the diagonal Green's function in Nambu space, while $F_i$, $F_j$, and $F_\gamma$ are components of the anomalous Green's function. Both $G$ and $\vec F$ depend on the superconducting order parameter $\boldsymbol{\Delta}(k)$; their explicit definitions are given in Appendix~\ref{app:eliashberg_feedback}.

Near the pairing onset temperature we assume that  the superconducting gap $\Delta$ is small compared to the Fermi energy $E_F$, even when the pairing interaction itself is not weak. This permits a Ginzburg--Landau-type expansion\footnote{
Notice that the expansion here is in terms of the magnitude of the superconducting order parameter $\Delta$. 
For spin-triplet $p$-wave pairing, the order parameter can be written as $\boldsymbol{\Delta}(k)\cdot \boldsymbol{\sigma} i \sigma_2$,
where $\sigma_i$ are the spin Pauli matrices (see the off-diagonal term in Eq.~\eqref{eq:McMillanSigma_app2}). 
To determine the relative stability of different spin-triplet pairing states, one would need a more refined Ginzburg--Landau expansion involving quartic combinations of $\Delta_\alpha^* \Delta_\beta \Delta_\gamma^* \Delta_\delta$ with several independent coefficients~\cite{Mermin1973,Vollhardt2013}. 
That is not our goal here; for the present discussion, the simpler expansion in Eq.~\eqref{eq:GL} is sufficient.
}
of the thermodynamic potential,
\begin{align}
\Omega_{\LW} =  \Omega_n+ a \bigg{|}\frac{\Delta}{E_F}\bigg{|}^2 + b \bigg{|}\frac{\Delta}{E_F}\bigg{|}^4+\cdots .
\label{eq:GL}
\end{align}
The quadratic coefficient $a$ is determined by the linearized gap equation and therefore does not yet 
include the feedback of the superconducting state on the interaction that mediates pairing. That feedback first enters at quartic order, through the coefficient $b$, which controls the stability of the ordered state and contributes directly to the condensation energy~\cite{Tsoncheva2005,Haslinger2003a,Haslinger2003,Chakravarty2003}. 

To isolate this effect, we write
\begin{align}
b = b^{\text{w.c.}}+\delta b ,
\end{align}
where $b^{\text{w.c.}}$ is the usual weak-coupling contribution and $\delta b$ is the additional term generated by the feedback of superconductivity on the interaction itself. The corresponding condensation energy is
\begin{equation}  \label{eq:condenergy}
\left[\Omega_{\LW}\right]_{\text{cond}}
= -\frac{a^2}{4 b}.
\end{equation}
Hence a negative $\delta b$ increases the magnitude of the condensation-energy gain and further stabilizes the superconducting state.

Expanding $\Phi_{\RPA}$ to second order in the superconductivity-induced correction $\delta\hat{\chi}^{(2)}$, one finds~\cite{Brinkman1974,Armitage1976}
\begin{align}
\delta \Phi_{\RPA}^{(4)}
&= \delta b \bigg| \frac{\Delta}{E_F} \bigg|^4 
= -\Tr\!\left[\left(
-\frac{I}{2}\big[1-I\hat{\chi}^{(0)}\big]^{-1}
\delta\hat{\chi}^{(2)}
\right)^2\right].
\label{eq:PhiRPA4}
\end{align}
Here we used Eqs.~\eqref{eq:LW1} and \eqref{eq:PhiRPA}, together with
\begin{align}
\chi_{ij}= \chi_{ij}^{(0)} + \delta \chi_{ij}^{(2)} + \mathcal{O}\bigg( \bigg(\frac{|\Delta|}{E_F}\bigg)^4 \bigg),
\end{align}
where the superscript ``$(n)$'' denotes order $n$ in $\Delta$. The term linear in $\hat{\chi}-\hat{\chi}^{(0)}$ contributes only\footnote{See Ref.~\onlinecite{Armitage1976}, Chapter 8 by P. W. Anderson and W. F. Brinkman, pp.~374--375.}
to the conventional weak-coupling coefficients $a|\Delta|^2$ and $b^{\text{w.c.}}|\Delta|^4$, but not to the genuine feedback contribution $\delta b |\Delta|^4$.

Equation~\eqref{eq:PhiRPA4} is the central result of this subsection. Within this RPA-like framework, it shows that superconducting feedback produces an additional condensation-energy gain by modifying the same fluctuations that mediate pairing.\footnote{Strictly speaking, the condensation energy corresponds to the \textit{stationary-point value} of the functional $\Omega_{\LW}-\Omega_n$ and is given by Eq.~\eqref{eq:condenergy} when the expansion in Eq.~\eqref{eq:GL} is truncated at quartic order in $\Delta$. These distinctions do not affect the main conclusion: feedback leads to a universal additional condensation-energy gain.}

The free-energy approach both exposes this energy gain and reproduces the familiar strong-coupling gap equations. The self-consistent gap equations and the feedback correction follow from the same $\Phi_{\RPA}$ functional. As shown in Appendices~\ref{app:eliashberg_feedback} and \ref{app:mcmillan}, the generalized Eliashberg equations emerge from this common starting point, placing the superconducting self-energies and the feedback-induced modification of the mediating fluctuations within a single variational framework.

Two features of Eq.~\eqref{eq:PhiRPA4} are especially important:
\begin{enumerate}[label={(\arabic*)}]
\item
At Matsubara frequencies, both $\hat{\chi}$ and $\hat{\Gamma}$ are real symmetric matrices. This ensures that $\Tr[(\hat{\Gamma}^{(0)}\delta\hat{\chi}^{(2)})^2] > 0$.

\item
The magnitude of $\delta b$ depends not only on the size of $\hat{\Gamma}^{(0)}\equiv-(I/2)[1-I\hat{\chi}^{(0)}]^{-1}$ and $\delta\hat{\chi}^{(2)}/|\Delta|^2$, but also on how strongly their weight overlaps in frequency--momentum space.
\end{enumerate}

Although the size of $\delta b$ is system dependent, its sign is fixed within this stable RPA regime: $\delta b < 0$. A negative $\delta b$ may be viewed as a reduction of the effective repulsion between Cooper pairs, or equivalently as an additional attraction generated by feedback. The result is an enhancement of the equilibrium gap amplitude and of the overall stability of the superconducting state. In magnetically mediated superconductors this effect is especially consequential, since the relevant pairing channels are often non-$s$-wave and therefore intrinsically more fragile.

\subsection{Constructive feedback across distinct electronically mediated pairing channels}
\label{sec:universal}

Eq.~\eqref{eq:PhiRPA4} reveals a broader energetic mechanism that is not confined to the paramagnon-triplet example considered above. What matters is not the microscopic identity of the fluctuations, but the existence of an RPA-like free-energy structure such as Eq.~\eqref{eq:PhiRPA}. Once that condition is met, superconductivity necessarily reshapes the same fluctuating state that contributes to pairing, yielding an additional condensation-energy gain. We now show that this constructive feedback reappears across several distinct realizations of electronically mediated pairing, including antiferromagnetic singlet, ferromagnetic triplet, and multiorbital settings.

\subsubsection{Antiferromagnetic fluctuations and spin-singlet $d$-wave pairing}
\label{sec:AFM}

As a first illustration of the broader scope of Eq.~\eqref{eq:PhiRPA4}, we consider spin-singlet $d$-wave pairing mediated by antiferromagnetic fluctuations. 
A natural setting is the square-lattice Hubbard model near half filling, relevant to the cuprates. 
In this regime, near nesting of the Fermi surface gives rise to strong antiferromagnetic spin fluctuations~\cite{Monthoux1994}, 
so that $\chi(i\Omega_m,\vecq)$ is sharply peaked near $\vecq=(\pi,\pi)$, in units of the inverse lattice constant. 
These fluctuations then dominate the Luttinger--Ward functional.

We assume that their contribution to the free energy is again described by the same $\Phi_{\mathrm{RPA}}$ as in Eq.~\eqref{eq:PhiRPA}, 
except that (i) the spin-susceptibility tensor $\chi_{ij}$ is isotropic, and (ii) the form of $\chi_{ij}$ in Eq.~\eqref{eq:chisc} is replaced by its spin-singlet counterpart
~\footnote{
In the normal state ($T>T_c$), the Luttinger--Ward plus RPA construction for magnetic fluctuations gives
$\Gamma^{(\mathrm{ph})}(q)\equiv-\delta \Phi_{\mathrm{RPA}}/\delta \chi(q)$.
Substituting the RPA particle-hole vertex,
$\Gamma^{(\mathrm{ph})}(q)=(3/2) I/[1-I\chi(q)]$ and integrating over $\chi(q)$ yields Eq.~\eqref{eq:PhiRPA}.
Apart from numerical prefactors in the terms first and second order in $I$, this is the magnetic-fluctuation component of the FLEX vertex; see, e.g., Ref.~\onlinecite{Bickers1989}.
These low-order prefactors are unimportant for our discussion because, near the disappearance of magnetic order, $\Gamma^{(\mathrm{ph})}(q)$ is dominated by higher-order terms.
For $T<T_c$, we assume that the dominant part of $\Phi_{\mathrm{RPA}}$ retains the same form, with $\chi$ replaced by the superconducting-state susceptibility.
}
Since the structure of $\Phi_{\mathrm{RPA}}$ is unchanged, the feedback term in Eq.~\eqref{eq:PhiRPA4} follows immediately.

In the antiferromagnetic case, the constructive feedback also admits a simple microscopic interpretation, as revealed by detailed Eliashberg analyses within the FLEX approximation~\cite{Monthoux1994,Scalapino2012}. 
These analyses show that the antiferromagnetic peak of $\chi(i\Omega_m,\vecq)$ near $\vecq=(\pi,\pi)$ is only weakly affected by the onset of superconductivity, 
consistent with the expectation that the superconducting gap mainly suppresses $\chi(i\Omega_m,\vecq)$ at low bosonic frequencies and small momentum transfers.  
The key reason for the constructive feedback is that low-frequency magnetic fluctuations tend to be pair breaking~\cite{Millis1988}, 
whereas the fluctuations most effective for pairing lie at higher frequencies~\cite{Monthoux1994a,Scalapino2012}. 
Suppressing primarily the low-frequency magnetic fluctuation therefore reduces pair breaking and enhances the superconducting gap, 
thereby giving rise to constructive feedback.

\subsubsection{Ferromagnetic paramagnon fluctuations and spin-triplet pairing}
\label{sec:ferromagnetic}

The reduction of pair breaking discussed above is one microscopic route to the constructive feedback encoded in Eq.~\eqref{eq:PhiRPA4}. 
A second route, more directly aligned with the central mechanism of this paper, arises in superconductivity mediated by ferromagnetic paramagnons, as discussed in Sec.~\ref{sec:LW}. 
Equation~\eqref{eq:PhiRPA4} already shows that the paramagnon-mediated attraction is generically enhanced once superconductivity develops, independent of the detailed spin structure of the spin-triplet order parameter.
We now illustrate the microscopic origin of this effect for a particular spin-triplet pairing state.

As a concrete example, we consider the Anderson--Brinkman--Morel pairing state, with spin structure
\(
(\uparrow \downarrow + \downarrow \uparrow)/\sqrt{2}
\).
The effective pairing interaction entering the BCS gap equation can be obtained from the second functional derivative of $\Phi_{\mathrm{RPA}}$ in Eq.~\eqref{eq:PhiRPA} 
with respect to the anomalous Green's functions of the ABM state;  see Eq.~\eqref{eq:Eliashberg_app2b} of Appendix~\ref{app:eliashberg_feedback}. 
One finds~\cite{Brinkman1974}
\[
I_{\mathrm{eff},\uparrow\downarrow}(q)
=
\frac{1}{2}
\left[
\frac{I}{1-I\chi_{zz}(q)}
-
\frac{2I}{1-I\chi_{xx}(q)}
\right],
\]
where the factor of $2$ in the second term reflects the degeneracy of $\chi_{xx}$ and $\chi_{yy}$ due to the residual rotational symmetry in the $xy$ plane of spin space. 
The first term, arising from longitudinal spin fluctuations, is repulsive, 
whereas the second term, arising from transverse spin fluctuations, is attractive.
\footnote{The denominators $1-I\chi_{zz}(q)$ and $1-I\chi_{xx}(q)$ are both positive because the normal state is assumed to remain paramagnetic, without magnetic long-range order.}

The ABM state is spin-anisotropic: the superconducting gap suppresses the longitudinal susceptibility $\chi_{zz}$,
predominantly at small $q=(i\Omega_m,\vecq)$, 
while leaving $\chi_{xx}$ and $\chi_{yy}$ nearly unchanged. As a result, the repulsive part of $I_{\mathrm{eff},\uparrow\downarrow}(q)$ is reduced, 
whereas the attractive part is essentially unaffected. The net pairing attraction is therefore enhanced. 
In this case, the constructive feedback originates from the spin anisotropy of the pairing interaction.

The two microscopic routes discussed here and in Sec.~\ref{sec:AFM}---reduction of pair breaking 
and spin-anisotropy-driven reduction of repulsion---thus arise from the same RPA form of the Luttinger--Ward functional, $\Phi_{\mathrm{RPA}}$, 
and are both naturally contained in Eq.~\eqref{eq:PhiRPA4}. 
The conclusion that the feedback correction is constructive therefore applies broadly to both singlet and triplet pairing, 
and to magnetic fluctuations with very different structures in $\chi(q)$.

\subsubsection{Pairing mediated by magnetic fluctuations in the multiorbital Kanamori--Hubbard model}
\label{sec:multiorbital}

The scope of the constructive feedback mechanism becomes even clearer in the Kanamori--Hubbard model, which generalizes the on-site Hubbard interaction to a multiorbital setting. This example shows that the energetic principle identified here remains operative in a richer interaction setting beyond the single-orbital model. This broader scope is especially relevant for many modern unconventional superconductors, where multiorbital physics often plays an essential role.

As an example, we consider a Kanamori--Hubbard model describing the on-site interactions among $t_{2g}$ $d$-electrons, with interaction Hamiltonian~\cite{Takimoto2004,Witt2021,Kuroki2009,Yada2005,Georges2013},
\begin{equation}
H_{\mathrm{int}}
=
\frac{1}{4}
\sum_i
\sum_{\xi_1 \xi_2 \xi_3 \xi_4}
\Gamma^0_{\xi_1 \xi_4,\xi_3 \xi_2}
\,c^\dagger_{i\xi_1} c^\dagger_{i\xi_2} c_{i\xi_3} c_{i\xi_4}.
\end{equation}
Here, $c_{i\xi}^\dagger$ creates an electron at site $i$ in the state $\xi=(\ell,\sigma)$, where $\ell\in\{d_{xz},d_{yz},d_{xy}\}$
 labels the $t_{2g}$ orbitals of a $d$-electron and $\sigma\in\{\uparrow,\downarrow\}$ the spin.

Assuming SU(2) spin symmetry, the bare interaction vertex can be written as
\begin{align}
\Gamma^0_{\xi_1 \xi_4,\xi_3 \xi_2}
=
&-\frac{1}{2}
U^s_{\ell_1 \ell_4,\ell_3 \ell_2}\,
\boldsymbol{\sigma}_{\sigma_1 \sigma_4}
\!\cdot\!
\boldsymbol{\sigma}_{\sigma_2 \sigma_3}
\nonumber\\
&+
\frac{1}{2}
U^c_{\ell_1 \ell_4,\ell_3 \ell_2}\,
\delta_{\sigma_1 \sigma_4}\delta_{\sigma_2 \sigma_3}.
\end{align}
The nonvanishing interaction matrix elements are
\begin{equation}
U^{\nu}_{\ell_1 \ell_4,\ell_3 \ell_2}
=
\begin{cases}
U          & \text{if } \ell_1=\ell_4=\ell_3=\ell_2,\\
\alpha_\nu & \text{if } \ell_1=\ell_3\neq \ell_2=\ell_4,\\
\beta_\nu  & \text{if } \ell_1=\ell_4\neq \ell_2=\ell_3,\\
J'         & \text{if } \ell_1=\ell_2\neq \ell_3=\ell_4,
\end{cases}
\quad
\nu\in\{s,c\},
\nonumber
\end{equation}
with
\[
\alpha_s=U',        \quad       \beta_s=J,      \quad    \alpha_c=-U'+2J, \quad \beta_c=2U'-J.
\]
Here, $U$ and $U'$ denote the intra-orbital and inter-orbital Hubbard repulsions, respectively, 
while $J$ and $J'$ are the Hund's coupling and pair-hopping interaction, respectively~\cite{Georges2013}.
If one assumes the additional SO(3) orbital rotational invariance, then $U'=U-2J$ and $J'=J$. 

The matrices $\hat{U}^s$ and $\hat{U}^c$ generate effective spin and charge interactions through exchange of the corresponding collective fluctuations. 
Near a magnetic instability, spin fluctuations dominate and charge fluctuations may be neglected. 
In this regime, the Luttinger--Ward grand potential $\Omega$ can again be written in the form of Eq.~\eqref{eq:LW1}, with the LW functional approximated by
\begin{align}
\Phi_{\mathrm{RPA}}
=
\frac{1}{2}\Tr\!\left[\ln\!\left(1-\hat{U}^s \hat{\chi}^s\right)\right].
\end{align}
Compared with Eq.~\eqref{eq:PhiRPA}, the trace now runs over both spin and orbital degrees of freedom, 
while $\hat{U}^s$ is itself a matrix in orbital space. For spin-singlet pairing, this RPA form leads to the effective spin-fluctuation-mediated pairing interaction
\begin{equation}
\hat{\Gamma}(q)
=
\frac{3}{2}\left(1-\hat{U}^s \hat{\chi}^s(q)\right)^{-1}\hat{U}^s,
\end{equation}
which is a scalar in spin space but a matrix in orbital space. Here $\hat{\chi}^s(q)$ is the multiorbital generalization of Eq.~\eqref{eq:chisc}; 
its detailed orbital structure is not needed for the present argument.
\footnote{As in the single-orbital case, Taylor expansion of $\hat{\Gamma}$ in powers of $\hat{U}^s$ overcounts the first- and second-order contributions. 
This is unimportant here because, near a magnetic instability, the interaction is dominated by higher-order terms.}

Expanding the spin-susceptibility matrix as
\begin{equation}
\hat{\chi}^s(q)
=
\hat{\chi}^{s,(0)}(q)
+
\delta \hat{\chi}^{s,(2)}(q)
+
\mathcal{O}\!\left(|\Delta|^4/E_F^4\right),
\nonumber
\end{equation}
and then expanding $\Phi_{\mathrm{RPA}}$ to second order in $\delta \hat{\chi}^{s,(2)}$, exactly as in Sec.~\ref{sec:LW}, we obtain
\begin{equation}
\delta \Phi_{\mathrm{RPA}}^{(4)}
=
-\Tr\!\left[
\left(
\frac{1}{2}
\left(1-\hat{U}^s \hat{\chi}^{s,(0)}\right)^{-1}
\hat{U}^s
\delta \hat{\chi}^{s,(2)}
\right)^2
\right]
< 0.
\end{equation}
At Matsubara frequencies, both $\hat{\chi}^s$ and $\hat{\Gamma}$ are real symmetric matrices, so the trace of the matrix square is positive. 
Thus, even in the multiorbital case, the onset of superconductivity produces a constructive feedback on the magnetic-fluctuation-mediated pairing interaction.

These examples establish that the incentive to pair when magnetism or other ordering disappears
(which we call ``constructive feedback") is not confined to a particular microscopic 
ordering, but reflects a broader energetic principle of electronically mediated superconductivity. 
Earlier literature developed related feedback formalisms in response to different
physical questions, most notably phase selection among nearly degenerate superconducting
states~\cite{Brinkman1974,Amin2020,Kozii2019} and Eliashberg analyses of magnetically
mediated pairing~\cite{Monthoux1994}. Here we show that these seemingly distinct cases
help answer a central question for unconventional superconductors: why superconductivity
tends to occur in close proximity to an ordered state (ferro- or antiferromagnetic,
etc.) that is on the verge of disappearing.

Moreover, physically we understand the origin of this effect.
When nearby order becomes attenuated or cannot fully develop, the associated fluctuating state  carries an excess free-energy cost. The opening of a pairing gap
can lower that cost and thereby provide an additional stabilization of the paired phase.
This, in turn, provides an understanding of the classic non-QCP phase diagram in which
the pairing gap is largest near the endpoint for ordering.

\section{Understanding the Non-QCP Phase Diagram: Contemporary Superconductors Near Disappearing Order}
\label{sec:strongpairing}

\begin{figure*}
\includegraphics[width=0.98\linewidth, clip]
{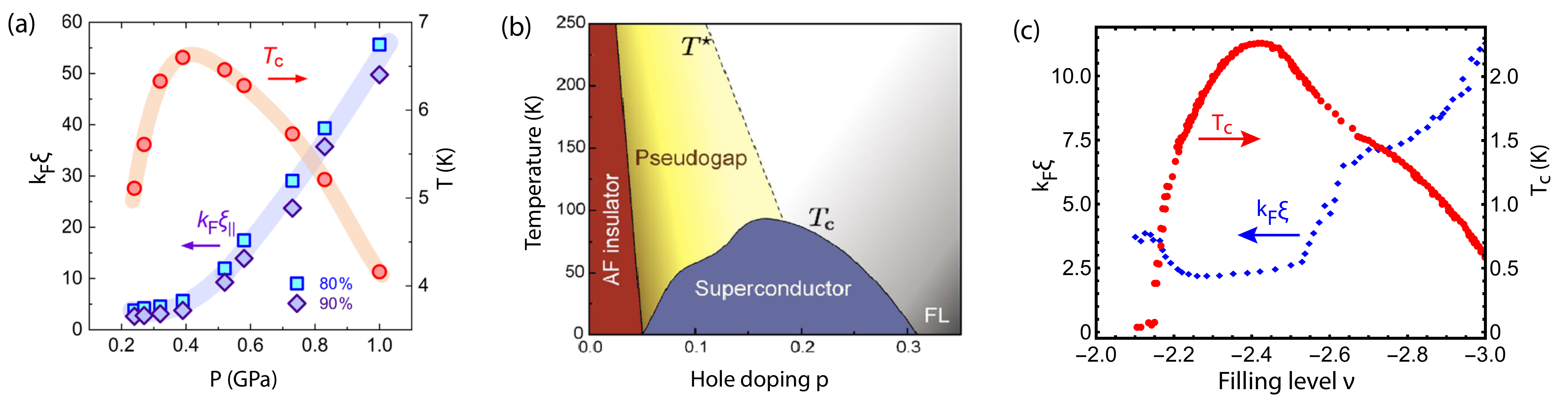}
\caption{Examples of superconductors that exhibit non-QCP-like phase diagrams, with superconducting domes near an insulating phase.
(a) Experimental dimensionless Ginzburg--Landau coherence length $k_F\xi^{\mathrm{GL}}$ and transition temperature $T_c$ versus pressure $P$ for the organic superconductor
$\kappa$-(BEDT-TTF)$_4$Hg$_{2.89}$Br$_8$, taken from Ref.~\onlinecite{Suzuki2022}.
(b) Representative experimental temperature-versus-hole-doping phase diagram for hole-doped cuprates, taken from Ref.~\onlinecite{Vignolle2013}.
(c) Experimental dimensionless Ginzburg--Landau coherence length $k_F\xi^{\mathrm{GL}}$ and transition temperature $T_c$ versus filling factor $\nu$ per moir\'e unit cell in twisted trilayer graphene, extracted from Ref.~\onlinecite{Park2021}.
In all three panels, the leftmost regime corresponds to a correlated phase. Small values of $k_F\xi^{\mathrm{GL}}$ in (a) and (c), and the high pseudogap onset temperature $T^*$ in (b), are used here as signatures of strong pairing.
In panel (c), insulating states occur on both sides of the superconducting dome, at $\nu=-2$ and near $\nu=-3$.
}
\label{fig:Fig4}
\end{figure*}

The non-QCP-like phase diagram illustrated schematically in Fig.~\ref{fig:Fig1b}(b) is not an isolated case, but recurs in several unconventional superconductors. Figure~\ref{fig:Fig4} compares three systems with very different microscopic properties that nevertheless exhibit superconducting domes adjacent to correlated insulating phases. This recurring structure provides a natural setting in which to apply the energetic perspective developed above.

In many recently discovered superconductors, including moir\'e materials such as twisted graphene, superconductivity appears next to correlated or partially ordered phases even though the existence, or relevance, of a sharp quantum critical point remains unsettled. Because the Luttinger--Ward--RPA treatment developed above is generic rather than material specific, we use its energetic principle to organize this broader phenomenology. Our focus is the pairing scale, as reflected in the gap $\Delta(T)$, rather than a microscopic description of phase coherence and the resulting $T_c$. Distinguishing $\Delta(T)$ from $T_c$ is essential in the strong-pairing regime considered here.

The characteristic dome-shaped dependence of the transition temperature $T_c$
in Fig.~\ref{fig:Fig1b}(b)
provides a natural setting for comparing the interplay between pairing and magnetism in both the QCP and non-QCP scenarios. In the conventional QCP picture, superconductivity is often regarded as strongest near the peak of the dome, where $T_c$ reaches its maximum. In the broader class of non-QCP-like phase diagrams considered here, however, an important distinction emerges: the regime of strongest pairing need not coincide with the regime of highest $T_c$. Instead, the strongest pairing is expected at the edge of the dome closest to the parent ordered phase, as illustrated in Fig.~\ref{fig:Fig1b}(b), even though $T_c$ there may already be reduced.

Although this second scenario frequently involves magnetism in an insulating parent phase and is therefore more closely tied to Mott physics than to the itinerant-electron setting considered above, the underlying energetic principle remains. In both cases, pairing emerges close to a suppressed magnetic or correlated phase. The attenuation of that nearby phase can then play a constructive, rather than purely competitive, role in stabilizing superconductivity. 

It is therefore natural to ask how the dome-shaped dependence of $T_c$ arises in this class of superconductors. On the weak-coupling side, far from the parent ordered phase, $T_c$ is small because the relevant magnetic or particle-hole fluctuations are too weak to generate strong pairing. As the system moves toward the ordered regime, these fluctuations become stronger, enhancing the pairing interaction and causing $T_c$ to rise.

By continuing this trend toward the ordered side of the phase diagram, one is naturally led to the regime of strongest pairing. Stronger pairing, however, does not necessarily imply a higher transition temperature. As the interaction continues to increase, the system eventually enters a strong-pairing regime in which the mobility of the pairs becomes reduced. This is a familiar feature of the BCS--BEC crossover~\cite{Nozieres1985,Chen2024}: stronger attraction promotes pair formation, but the resulting composite pairs become less mobile, so the onset of long-range phase coherence is suppressed to lower temperature. The pair hopping that controls this mobility ultimately derives from the hopping of the constituent fermions and becomes strongly constrained when the binding interaction is large. As a result, a dome-shaped dependence of $T_c$ can emerge, with $T_c$ becoming small in both the weak-pairing and strong-pairing limits.\footnote{
We note that to describe quantitatively the behavior of $T_c$ on the strong-coupling side of this dome, one needs to go beyond the formalism developed in the previous sections.}

The non-QCP scenario in Fig.~\ref{fig:Fig1b}(b) is therefore not a departure from the general mechanism discussed above, but one of its most natural consequences. The same nearby correlated state or ordering tendency that underlies the pairing interaction also creates the conditions under which superconductivity gains an additional energetic advantage through feedback on the associated fluctuations. What changes across the dome is not simply the existence of pairing glue, but the balance between pairing strength and pair mobility.

\subsection{Three Representative Material Classes: Moir\'e Graphene, Organic, and Cuprate Superconductors}

We now turn to the representative materials shown in Fig.~\ref{fig:Fig4}. Despite their different microscopic details, they share a common phase-diagram structure: the strongest signatures of pairing occur adjacent to an insulating, and in some cases ordered, phase rather than at the maximum of $T_c$.

For the organic~\cite{Suzuki2022} and cuprate materials shown in Figs.~\ref{fig:Fig4}(a) and \ref{fig:Fig4}(b), the neighboring insulator has a Mott origin; in the cuprates, it also exhibits clear antiferromagnetic order. 
In the organic and moir\'e graphene cases, however, the ordering tendencies of the adjacent insulating phases are not yet fully resolved. For the organic superconductor $\kappa$-(BEDT-TTF)$_4$Hg$_{2.89}$Br$_8$, the nearby insulating regime has been argued to possess a spin-liquid-like character inherited from the doped Mott state. In the moir\'e systems, the nature of the proximate order remains under active debate. Existing evidence points to broken flavor symmetry, but the dominant order parameter---whether primarily spin, valley, or mixed spin-valley in character---has not been uniquely identified.

From the perspective of the present paper, this microscopic uncertainty does not undermine the broader phase-diagram lesson. In all of these cases, superconductivity develops near the weakening of a correlated phase or incipient ordering tendency and may therefore gain condensation energy by relieving the free-energy cost of the nearby fluctuating state, even when no long-range order is firmly established. The proximate state need not be a fully identified magnet; what matters is a strong but incompletely realized correlated tendency. Related examples include CeCoIn$_5$, PuCoGa$_5$~\cite{Sidorov2002,Ramshaw2015}, and \heliumthree, where the relevant nearby order is only partially or indirectly visible in the phase diagram.

A recurring experimental trend also bears on this interpretation. Enhancing screening, and thereby reducing the effective Coulomb interaction, can weaken nearby insulating states while strengthening superconductivity. In moir\'e graphene, such trends have been invoked both against~\cite{Barrier2024,Gao2024} and in favor of~\cite{Saito2021,Stepanov2020,Liu2021} phononic pairing mechanisms. Related observations occur in heavy-fermion materials, where tuning that weakens magnetic order can promote superconductivity~\cite{Landaeta2018}, even though these systems are not generally regarded as natural candidates for phonon-mediated pairing. Such experiments therefore do not provide decisive evidence against an electronic mechanism; they are also consistent with the present picture, in which superconductivity tracks the retreating boundary of a correlated phase.

These systems also exhibit signatures of strong pairing~\cite{Chen2024}. Although such signatures do not by themselves identify the pairing mediator, they are consistent with an electronically driven mechanism. One useful indicator is a small dimensionless Ginzburg--Landau coherence length $k_F\xi^{\mathrm{GL}}$~\cite{Park2021}. Figure~\ref{fig:Fig4}(a) illustrates the distinction between the pairing and coherence scales particularly clearly for the organic superconductor: the minimum of $k_F\xi^{\mathrm{GL}}$, used here as an indicator of the strongest pairing tendency~\cite{Chen2024}, lies closer to the insulating boundary than the maximum of $T_c$.

More generally, for all three systems shown in Fig.~\ref{fig:Fig4}, the strongest pairing signatures appear closest to the neighboring insulating phase. In the cuprate case [Fig.~\ref{fig:Fig4}(b)], this is reflected in the high pseudogap onset temperature $T^*$, which is used here as a proxy for the pairing scale. A similar trend appears in twisted trilayer graphene [Fig.~\ref{fig:Fig4}(c)], where the smallest $k_F\xi^{\mathrm{GL}}$ occurs near the correlated insulator\footnote{
In this system the dimensionless coherence length shows a slight upturn near the insulating phase, although the experimental uncertainties are large. Recent tunneling measurements~\cite{Park2026} also indicate that the excitation gap does not peak precisely where $k_F\xi^{\mathrm{GL}}$ is smallest.
}
at moir\'e unit cell filling $\nu=-2$.

These materials therefore illustrate the distinction emphasized throughout this paper: the strongest pairing signatures occur not at the top of the dome, but at the edge nearest the neighboring correlated phase. In that regime, superconductivity lies closest to the fluctuations associated with the parent state and can acquire a larger pairing scale, while reduced pair mobility suppresses long-range phase coherence and lowers $T_c$.

Unlike the other two systems, twisted graphene shows evidence that the superconducting phase itself breaks translational symmetry through a form of Kekul\'e ordering distinct from that observed in the insulating state~\cite{Nuckolls2024}. Experiments further indicate that the Kekul\'e order remains strong where superconductivity is most robust~\cite{Kim2023}. This persistence argues against a simple scenario in which the weakening of Kekul\'e order supplies the dominant pairing fluctuations, although it does not exclude a contribution from those fluctuations. More generally, when an order parameter remains robust or strengthens where pairing is strongest, one should distinguish carefully between an order that drives pairing and a broken symmetry that coexists with it.

Given our emphasis on the twisted graphene family, it is natural to ask why superconductivity so often appears in systems with flat electronic bands~\cite{Cao2018a,Cao2018b}. Several mechanisms may be relevant, and their relative importance may vary from material to material. One frequently cited explanation is the enhanced density of states associated with flat bands. However, this alone is unlikely to be the full story. In heavy-fermion superconductors, the large effective masses likewise produce a high density of states, yet the resulting transition temperatures are generally modest and there is only limited evidence for exceptionally strong pairing.\footnote{
An interesting exception is CeCoIn$_5$, which has the highest $T_c$ among Ce-based heavy-fermion superconductors and exhibits pseudogap-like strong-pairing behavior. Notably, it does not show magnetic order in zero field.
}

It is therefore useful to consider a more indirect possibility. 
A substantial literature shows that flat bands can strongly promote ferromagnetism and other interaction-driven symmetry breaking~\cite{Tasaki1998}. While the precise role of flat bands in enhancing superconductivity remains unsettled, part of their importance may lie in their tendency to enable nearby ordered states. 
Within the framework discussed here, such proximate ordering can provide the fluctuations that promote pairing while also supplying the additional energetic stabilization that accompanies the onset of superconductivity.

\section{Broader Implications and Conclusions}
\label{sec:conclusion}

A wide variety of unconventional superconductors appear in close proximity to magnetic or correlated insulating phases. At first sight this relationship seems paradoxical, since long-range magnetic order and superconductivity are often mutually exclusive. Yet across many material families~\cite{Weng2016}, including organics, cuprates, heavy fermions, and possibly moir\'e graphene systems, the emergence of superconductivity near the disappearance of a nearby ordered phase is a recurring and striking feature of the phase diagram. The central goal of this work has been to provide a more general explanation for this widespread phenomenology.
Figure~\ref{fig:Fig1b} illustrates two generic phase-diagram structures. Our focus has been on the non-QCP-like case in panel (b), where quantum-critical behavior is less apparent.

To address these phase diagrams, one must consider the energetics. The key ingredient we have focused on here is that when a magnetic or related particle-hole instability is strong but cannot be fully realized, the paired state can gain an additional energy advantage through feedback effects that relieve part of the excess energy associated with the fluctuating state. In this sense, the proximity of superconductivity to disappearing order is not explained solely by the availability of strong pairing-mediating fluctuations; it also reflects a thermodynamic stabilization mechanism.
Within the RPA-like Luttinger--Ward framework developed in Sec.~\ref{sec:feedbackLW}, this mechanism takes a particularly transparent form: pairing is associated with
an energy gain when it renormalizes the same collective fluctuation contribution that helps mediate pairing. This energetic effect is quite general and does not rely on the presence of a magnetic QCP.

The framework developed here also helps determine which nearby ordered states are plausible candidates for the pairing glue. If a given order is responsible for the dominant pairing fluctuations, its weakening should accompany the development of superconductivity. By contrast, when an order remains robust, or even becomes more pronounced, as pairing strengthens, it is unlikely to represent the primary source of the pairing interaction, although its fluctuations may still contribute. This provides a useful way to distinguish between an order that drives pairing and a broken symmetry that coexists with it in systems with multiple intertwined broken symmetries.

A central message of this work is that the strongest pairing tendency need not coincide with the maximum $T_c$, but instead can occur on the side of the superconducting dome closest to a weakened correlated phase or incipient order. As illustrated across several classes of unconventional superconductors, this regime is characterized by a comparatively small $T_c$, while the pairing scale, reflected in the magnitudes of $T^*$ and the excitation gap $\Delta$, and in the small dimensionless Ginzburg--Landau coherence length $k_F\xi^{\mathrm{GL}}$, remains large. This separation of pairing strength and transition temperature is not emphasized in the simplest quantum-critical framework, where strong pairing is typically assumed to track high $T_c$. Instead, it points to an alternative phase-diagram structure in which pairing strength and phase coherence evolve differently. Viewed in this way, the recurring proximity of superconductivity to magnetism or other correlated order reflects an energetic contribution beyond a simple ``glue"-based picture and suggests a qualitative guide for materials discovery: systems with strong but readily suppressible correlated tendencies may be promising candidates.

\section{Acknowledgment }

We thank W. P. Halperin for helpful communications and R. Boyack for discussions and
acknowledge useful advice from A. Carrington and C. Proust.
Z. W. is supported by the Innovation Program for Quantum Science and Technology (Grant No. 2021ZD0301904).
We also acknowledge the University of Chicago's Research Computing Center for their support of this work.

\appendix
\numberwithin{equation}{section}
\numberwithin{figure}{section}

\section{Landau scattering amplitudes and pairing strength in \texorpdfstring{$^3$He}{3He}}
\label{app:He3_landau}

In this appendix we provide additional details for the discussion of superfluid $^3$He in Sec.~\ref{sec:TcHe3}. 
Our goal is to clarify how the effective interaction in the paramagnon description is connected both to the experimentally inferred Landau scattering amplitudes and to the pairing strength that determines the transition temperature.

We consider the $\ell$-th partial-wave Landau quasiparticle scattering amplitudes $A_\ell^{s/a}$ in the density ($s$) and spin ($a$) channels, defined for the effective interaction Hamiltonian $\hat{H}^{\rm eff}$ in Eq.~\eqref{eq:Heff}. 
If the paramagnon theory were exact, these amplitudes would be related to the Landau interaction parameters $F_\ell^{s/a}$ through
\begin{equation}
A_\ell^{s/a}=\frac{F_\ell^{s/a}}{1+F_\ell^{s/a}/(2\ell+1)}.
\end{equation}
In practice, however, the paramagnon treatment is only approximate, so it is more meaningful to compare the theory directly with the measured scattering amplitudes $A_\ell^{s/a}$ rather than with the inferred parameters $F_\ell^{s/a}$. This choice is also natural from a formal point of view: $A_\ell^{s/a}$ is defined from the $\vec q$-limit of the vertex function, namely from the physical forward-scattering amplitude of quasiparticles on the Fermi surface, where screening from repeated particle-hole rescattering is fully included. Moreover, it is $A_\ell^{s/a}$, rather than $F_\ell^{s/a}$, that is most directly connected to the two-particle irreducible interaction entering the Cooper-channel Bethe--Salpeter equation when pairing occurs within a narrow shell about the Fermi surface~\cite{Patton1975}.

\subsection{Vertex functions and Landau scattering amplitudes}

We begin with the general four-point vertex
\begin{equation}
\GammaVar(p_1,p_2;p_1+q,p_2-q)\equiv \GammaVar(p_1,p_2;q),
\end{equation}
which describes the scattering of two incoming particles with momenta $(p_1,p_2)$ into outgoing particles with momenta $(p_1+q,p_2-q)$.
For a spin-rotation-invariant system, this process can be decomposed into spin-triplet and spin-singlet channels. The corresponding vertices obtained from the effective interaction $\hat{H}^{\text{eff}}$ in Eq.~\eqref{eq:Heff} are~\cite{Levin1979a}
\begin{align} \label{eq:vertextriplet_app2}
\GammaVar_{\rm trip}(p_1,p_2;q)
&=
\langle p_1+q,p_2-q;\uparrow\uparrow|\hat{H}^{\rm eff}|p_1,p_2;\uparrow\uparrow\rangle
\nonumber\\
&=
-J(q)+J(q+p_1-p_2) \nonumber \\
& \quad +V(q)-V(q+p_1-p_2),
\end{align}
and
\begin{align} \label{eq:vertexsinglet_app2}
\GammaVar_{\rm sing}(p_1,p_2;q)
&=
\langle p_1+q,p_2-q;0,0|\hat{H}^{\rm eff}|p_1,p_2;0,0\rangle
\nonumber\\
&=
3J(q)+3J(q+p_1-p_2) \nonumber \\
&\quad +V(q)+V(q+p_1-p_2),
\end{align}
where
\begin{align}
|p_1,p_2;\uparrow\uparrow\rangle
&=
c^\dagger_{p_1\uparrow}c^\dagger_{p_2\uparrow}|0\rangle,\\
|p_1,p_2;0,0\rangle
&\equiv
\frac{
c^\dagger_{p_1\uparrow}c^\dagger_{p_2\downarrow}
-
c^\dagger_{p_1\downarrow}c^\dagger_{p_2\uparrow}
}{\sqrt{2}}|0\rangle .
\end{align}
Because $\hat{H}^{\rm eff}$ is spin-rotation invariant, any other basis state in the triplet manifold gives the same result.
We also assume that $J(q)$ and $V(q)$ already denote effective interactions obtained after summing the relevant classes of diagrams, so that no further loop corrections are required in defining $\GammaVar$.

To relate these vertices to the Landau scattering amplitudes, we place the external particles on the Fermi surface and set all external frequencies to zero. 
The resulting vertex then depends only on the angle $\theta$ between the incoming momenta $\vec p_1$ and $\vec p_2$, and on the angle $\phi$ between the planes spanned by the incoming and outgoing momenta. This corresponds to the limit $|\Omega|/(v_F|\vec q|)\ll 1$. We denote the vertex functions in this limit by $\GammaVar_{\rm trip/sing}^{\vec q}(\theta,\phi)$.

The corresponding Landau quasiparticle scattering amplitudes are~\cite{Levin1983}
\begin{subequations}\label{eq:vertexqlimit_app2}
\begin{align}
A^s(\theta,\phi)
&\equiv
\frac{N_F}{2Z}
\Big[
3\GammaVar_{\rm trip}^{\vec q}(\theta,\phi)
+
\GammaVar_{\rm sing}^{\vec q}(\theta,\phi)
\Big], \\
A^a(\theta,\phi)
&\equiv
\frac{N_F}{2Z}
\Big[
\GammaVar_{\rm trip}^{\vec q}(\theta,\phi)
-
\GammaVar_{\rm sing}^{\vec q}(\theta,\phi)
\Big],
\end{align}
\end{subequations}
where
\begin{equation}
Z=1+\lambda_Z
\end{equation}
is the quasiparticle frequency-renormalization factor. Within the momentum-independent self-energy approximation used here, it is identified with the effective-mass enhancement $m^*/m$.

Taking the additional forward-scattering limit $|\vec q|\to 0$ implies $\phi\to 0$, so that $A^{s/a}(\theta,0)$ depends only on the angle between the two incoming quasiparticle momenta. The Landau parameters $A_\ell^{s/a}$ are then defined through the Legendre expansion
\begin{equation}
A^{s/a}(\theta,0)=\sum_{\ell=0}^{\infty}A_\ell^{s/a}P_\ell(\cos\theta),
\end{equation}
with
\begin{equation}
A_\ell^{s/a}
=
\frac{2\ell+1}{2}
\int_0^\pi d\theta\,\sin\theta\,
A^{s/a}(\theta,0)P_\ell(\cos\theta).
\end{equation}
Using Eqs.~\eqref{eq:vertextriplet_app2}, \eqref{eq:vertexsinglet_app2}, and \eqref{eq:vertexqlimit_app2}, one finds~\cite{Levin1979a}
\begin{subequations}\label{eq:Aell1_app2}
\begin{align}
A_1^s
&=
\frac{N_F}{Z}(2\ell+1)(3J_\ell-V_\ell)\big|_{\ell=1}
=
\frac{N_F}{Z}(9J_1-3V_1), \\
A_1^a
&=
\frac{N_F}{Z}(2\ell+1)(-J_\ell-V_\ell)\big|_{\ell=1}
=
\frac{N_F}{Z}(-3J_1-3V_1),
\end{align}
\end{subequations}
where $J_\ell$ and $V_\ell$ are the corresponding Legendre components of the static effective interactions $J(q)$ and $V(q)$. In particular,
\begin{equation} \label{eq:Jell}
J_\ell  = \frac{1}{2} \int_{0}^{\pi}  d \theta  \;  \sin \theta \; J(\Omega =0, |\vecq|) P_\ell(\cos \theta ) ,
\end{equation}
so that
\begin{align}
J(\Omega =0, |\vecq|)  = \sum_{\ell=0}^\infty (2\ell +1)J_\ell P_\ell (\cos \theta).
\end{align}
An analogous expansion holds for $V_\ell$. Note that $J_\ell$ and $A_\ell^{s/a}$ are defined with different normalization conventions in their Legendre expansions, and therefore differ by factors of $(2\ell+1)$.

\subsection{Cooper-channel projection and pairing strength}

We now turn to the Cooper channel, obtained by setting $p_1=-p_2$. 
Assuming that pairing is confined to an energy shell about the Fermi surface that is small compared with $E_F$, we set all external frequencies to zero and impose
\begin{equation}
|\vec p_1|=|\vec p_2|=|\vec p_1+\vec q|=|\vec p_2-\vec q|=k_F.
\end{equation}
Under these conditions, the relevant pairing interaction is
\begin{equation}
\GammaVar^{\rm BCS}(\phi)\equiv \GammaVar^{\vec q}(\pi,\phi),
\end{equation}
which depends only on the angle $\phi$ between the incoming pair $\{\vec p_1,-\vec p_1\}$ and the outgoing pair $\{\vec p_2,-\vec p_2\}$. Its partial-wave components are defined by
\begin{equation}
\GammaVar_\ell^{\rm BCS}
=
\frac{1}{2}\int_0^\pi d\phi\,\sin\phi\,
\GammaVar^{\rm BCS}(\phi)P_\ell(\cos\phi).
\end{equation}

For $^3$He, the relevant instability is spin-triplet, orbital-$p$-wave pairing, \textit{i.e.} $\ell=1$. Using Eq.~\eqref{eq:vertextriplet_app2}, one obtains
\begin{align}
\GammaVar^{\rm BCS}(\phi)
&=
\GammaVar_{\rm trip}(\vec p_1,-\vec p_1;\vec p_3,-\vec p_3)
\nonumber\\
&=
-J(|\vec p_3-\vec p_1|)
+J(|\vec p_3+\vec p_1|)
\nonumber\\
&\quad
+V(|\vec p_3-\vec p_1|)
-V(|\vec p_3+\vec p_1|),
\end{align}
where $|\vec p_1|=|\vec p_3|=k_F$ and $\phi$ is the angle between $\vec p_1$ and $\vec p_3$. Projecting onto the $\ell=1$ channel then gives the effective pairing strength
\begin{equation}\label{eq:lambdaDelta2_app2}
\frac{\lambda_\Delta^{\ell=1}}{Z}
\equiv
-\frac{1}{Z}\frac{N_F\GammaVar_{\ell=1}^{\rm BCS}}{2}
=
\frac{N_F}{Z}(J_1-V_1).
\end{equation}
Here, $\GammaVar_{\ell=1}^{\rm BCS}<0$ corresponds to an attractive interaction, so that $\lambda_\Delta^{\ell=1}>0$. The factor of $1/2$ appears because $\GammaVar^{\rm BCS}$ is an antisymmetrized vertex, whereas the pairing kernel entering the conventional definition of $\lambda_\Delta$ is written without this antisymmetrization factor.

Comparing Eqs.~\eqref{eq:Aell1_app2} and \eqref{eq:lambdaDelta2_app2}, we see that the pairing strength and the Landau amplitudes are governed by the same effective interaction parameters, namely $J_1/Z$ and $V_1/Z$. Within the paramagnon picture, the exchange contribution $J_1$ dominates over $V_1$ near a magnetic instability. Neglecting $V_1$ then yields the approximate relation
\begin{equation}
\frac{\lambda_\Delta^{\ell=1}}{Z}\approx \frac{A_1^s}{9}.
\end{equation}
This is the relation used in Eq.~\eqref{eq:Tc} of the main text to express the exponent in the transition temperature in terms of the measured Landau parameters, and it underlies the construction of Fig.~\ref{fig:Fig2}(c).

\section{Self-consistent Eliashberg theory with feedback}
\label{app:eliashberg_feedback}

For electronically mediated pairing, the interaction kernel is not an externally prescribed bosonic propagator, but is instead generated by the same fermions that subsequently undergo pairing. A consistent treatment must therefore determine the pairing interaction and the superconducting state simultaneously.
This is also where a central difficulty has long remained.
In much of the literature, the attractive interaction is introduced phenomenologically, in close analogy with an external boson whose propagator is then modified by the opening of a fermionic gap. Here we show that, within a Luttinger--Ward formulation, feedback effects are not an additional ingredient imposed by hand, but instead emerge automatically once the same functional is used to determine both the fermionic self-energies and the fluctuation propagator.

Our starting point is the Luttinger--Ward functional $\Phi_{\RPA}$ shown in Fig.~\ref{fig:LW}. 
Unlike in conventional electron--phonon Eliashberg theory, where the phonon propagator $D$ is treated as an independent variational field~\cite{Eliashberg1963,Bardeen1964,Carbotte1990}, here $\hat{\chi}$ is not taken as independent. Instead, $\hat{\chi}$ is determined self-consistently from the fermionic Green's functions through Eq.~\eqref{eq:chisc}. Stationarity with respect to the fermionic self-energy then yields coupled equations for the diagonal and anomalous sectors. Although the resulting structure may appear natural in retrospect, deriving it from a single Luttinger--Ward functional is essential, because only then are the feedback effects entering the interaction kernel and those entering the fermionic self-energies treated on the same footing.

\subsection{General framework}

We write the Nambu self-energy as
\begin{align}\label{eq:Sigma3_app2}
\hat{\Sigma}(k)
=
\begin{pmatrix}
\Sigma(k) & \boldsymbol{\phi}(k)\cdot\bsigma\, i\sigma_2 \\
-i\sigma_2\,\boldsymbol{\phi}^*(k)\cdot\bsigma & -\Sigma(-k)
\end{pmatrix},
\end{align}
where $k=(i\omega_n,\vec k)$ and $\boldsymbol{\phi}(k)$ denotes the anomalous self-energy. Here $\boldsymbol{\sigma}=(\sigma_1,\sigma_2,\sigma_3)$ are the Pauli matrices in spin space. We reserve the notation $\boldsymbol{\Delta}$ for the gap function introduced later in the McMillan-level approximation. Both $\Sigma$ and $\boldsymbol{\phi}$ are treated here as variational quantities.
We restrict throughout to unitary triplet states, for which $\boldsymbol{\phi}^*(k)\times\boldsymbol{\phi}(k)=0$.

Stationarity of the Luttinger--Ward functional gives~\cite{Luttinger1960,Rainer1976}
\begin{equation}\label{eq:saddle_app2}
\frac{1}{2}\hat{\Sigma}^T(k)=\frac{\delta\Phi_{\RPA}}{\delta \hat G(k)}.
\end{equation}
Evaluating this functional derivative yields the coupled Eliashberg equations,
\begin{subequations}\label{eq:Eliashberg_app2}
\begin{align}
\Sigma(k)
&=
-\int_q \Tr[\hat{\Gamma}(q)]\,G(k+q), \\
\phi_i(k)
&=
\int_q \sum_{j=1}^3
\big[
\Tr[\hat{\Gamma}(q)]-\hat{\Gamma}(q)-\hat{\Gamma}^T(q)
\big]_{ij}
F_j(k+q).  \label{eq:Eliashberg_app2b}
\end{align}
\end{subequations}
The corresponding normal and anomalous propagators are
\begin{subequations}\label{eq:Gkdef_app2}
\begin{align}
G(k)
&=
\frac{i\omega_n-\Sigma(k)+\xi_{\vec k}}
{(i\omega_n-\Sigma(k))^2-\xi_{\vec k}^2-|\boldsymbol{\phi}(k)|^2}, \\
\vec F(k)
&=
\frac{\boldsymbol{\phi}(k)}
{(i\omega_n-\Sigma(k))^2-\xi_{\vec k}^2-|\boldsymbol{\phi}(k)|^2}.
\end{align}
\end{subequations}
Here we have assumed $\Sigma^T(-k)=\Sigma(-k)\approx -\Sigma(k)$ for the diagonal self-energy.

The interaction kernel is
\begin{equation}\label{eq:6.5_app2}
\hat{\Gamma}(q)=\frac{-I/2}{1-I\hat{\chi}(q)},
\end{equation}
which is a $3\times3$ matrix in spin space. Because $\hat{\chi}(q)$ is itself evaluated from the superconducting Green's functions, the interaction kernel already contains the feedback of pairing on the fluctuation spectrum. In this sense, Eqs.~\eqref{eq:Eliashberg_app2} and \eqref{eq:6.5_app2} do more than merely mimic the formal appearance of standard Eliashberg theory: they provide a closed and internally consistent formulation of electronically mediated pairing in which the same fermions both generate the pairing interaction and respond to the resulting order.

\subsection{Derivation of the Eliashberg equations}
\label{app:Eliashberg}

We now sketch the derivation of Eq.~\eqref{eq:Eliashberg_app2}. Starting from the stationarity condition in Eq.~\eqref{eq:saddle_app2}, it is convenient to treat the four Nambu blocks of $\hat{\Sigma}$ as independent matrix-valued variational variables:
\begin{subequations}
\begin{align}
\hat{\Sigma}(k)
&=
\begin{pmatrix}
\Sigma_{11}(k) & \Sigma_{12}(k) \\
\Sigma_{21}(k) & \Sigma_{22}(k)
\end{pmatrix}, \\
\hat{G}(k)
&=
\begin{pmatrix}
G_{11}(k) & G_{12}(k) \\
G_{21}(k) & G_{22}(k)
\end{pmatrix},
\end{align}
\end{subequations}
where each block remains a matrix in spin space.

In this notation, the $(i,j)$ component of the spin-susceptibility tensor $\hat{\chi}$ can be written as
\begin{widetext}
\begin{align}\label{eq:chisc2_app2}
\chi_{ij}(q)
=
-\frac14\int_p
\Tr\Bigg[
\begin{pmatrix}
\sigma_i & 0\\
0 & -\sigma_i^T
\end{pmatrix}
\begin{pmatrix}
G_{11}(p_-) & G_{12}(p_-)\\
G_{21}(p_-) & G_{22}(p_-)
\end{pmatrix}
\begin{pmatrix}
\sigma_j & 0\\
0 & -\sigma_j^T
\end{pmatrix}
\begin{pmatrix}
G_{11}(p_+) & G_{12}(p_+)\\
G_{21}(p_+) & G_{22}(p_+)
\end{pmatrix}
\Bigg],
\end{align}
\end{widetext}
with $p_\pm=p\pm q/2$. The prefactor $1/4$ arises from two distinct factors of $1/2$: one is inserted to avoid the usual double counting in Nambu space, and the other is fixed by the normalization convention for the spin operators.

Substituting the RPA functional $\Phi_{\RPA}$ from Eq.~\eqref{eq:PhiRPA} into the stationarity condition \eqref{eq:saddle_app2}, and then using the above expression for $\chi_{ij}(q)$, we obtain
\begin{widetext}
\begin{align}
\frac12
\begin{pmatrix}
\Sigma_{11}(k) & \Sigma_{12}(k)\\
\Sigma_{21}(k) & \Sigma_{22}(k)
\end{pmatrix}
=
\frac12\int_q\sum_{i,j=1}^3
\Bigg[\frac{-I}{1-I\hat\chi(q)}\Bigg]_{ij}
\Big(-\frac12\Big)
\begin{pmatrix}
\sigma_i & 0\\
0 & -\sigma_i^T
\end{pmatrix}
\hat G(k+q)
\begin{pmatrix}
\sigma_j & 0\\
0 & -\sigma_j^T
\end{pmatrix},
\end{align}
\end{widetext}
where we have used that the spin-susceptibility tensor $\hat\chi(q)$ is symmetric and an even function of $q$.

Comparing the corresponding Nambu blocks on the two sides of the above equation then yields
\begin{subequations}
\begin{align}
\Sigma_{11}(k)
&=
-\frac12\int_q\sum_{i,j=1}^3
\Bigg[\frac{-I}{1-I\hat\chi(q)}\Bigg]_{ij}
\sigma_i\,G_{11}(k+q)\,\sigma_j, \\
\Sigma_{12}(k)
&=
\frac12\int_q\sum_{i,j=1}^3
\Bigg[\frac{-I}{1-I\hat\chi(q)}\Bigg]_{ij}
\sigma_i\,G_{12}(k+q)\,\sigma_j^T.
\end{align}
\end{subequations}

If the diagonal contribution remains spin isotropic, then $G_{11}$ and $\Sigma_{11}$ are proportional to the identity in spin space, and the first equation reduces to
\begin{equation}\label{eq:Sigma11_app2}
\Sigma_{11}(k)
=
-\frac12\int_q \Tr\Bigg[\frac{-I}{1-I\hat\chi(q)}\Bigg] G_{11}(k+q).
\end{equation}
For the anomalous contribution, we write
\begin{equation}
\Sigma_{12}(k)=\boldsymbol{\phi}(k)\cdot\bsigma\, i\sigma_2,
\end{equation}
and project onto the components $\phi_i$. This yields
\begin{align}
\phi_i(k)
&=
\frac12\int_q \sum_{m,n,\ell=1}^3
\Bigg[\frac{-I}{1-I\hat\chi(q)}\Bigg]_{mn} 
 F_\ell(k+q) \nonumber \\
& \times \Big(
\delta_{i\ell}\delta_{mn}
-\delta_{im}\delta_{\ell n}
-\delta_{in}\delta_{m\ell}
\Big).
\label{eq:phii_app2}
\end{align}
Upon identifying $\Sigma_{11}=\Sigma$ and $G_{11}=G$, Eqs.~\eqref{eq:Sigma11_app2} and \eqref{eq:phii_app2} reduce directly to Eq.~\eqref{eq:Eliashberg_app2}.

More elaborate diagrammatic approximations to the free energy than the $\Phi_{\RPA}$ assumed here have also been considered in the $^3$He literature, most notably by Rainer and Serene~\cite{Rainer1976}. In such approaches, the stationarity conditions need not reduce to the compact Eliashberg form of Eq.~\eqref{eq:Eliashberg_app2}, and it is correspondingly less transparent how the feedback effects contained in the free energy are organized within the resulting gap equations. By contrast, the virtue of the present formulation is precisely that both the self-energy equations and the associated free-energy corrections follow from the same $\Phi_{\RPA}$ functional. In this sense, the present analysis provides a particularly simple and coherent route for incorporating feedback effects into an Eliashberg framework for electronically mediated pairing.

\section{McMillan-level approximation}
\label{app:mcmillan}

The full Eliashberg theory is formally appealing, but difficult to solve in practice because the diagonal and anomalous self-energies must be determined self-consistently together with the susceptibility. It is therefore useful to consider a simpler approximation in which the diagonal self-energy is retained only at its normal-state value, in the spirit of McMillan's treatment of strong-coupling superconductivity. This approximation is also close in spirit to the original Brinkman--Serene--Anderson analysis~\cite{Brinkman1974} of $^3$He. At the same time, deriving it directly from the same Luttinger--Ward functional is valuable because it makes explicit that feedback effects survive even after self-consistency in the diagonal contribution has been relaxed.

\subsection{Approximate gap equation}
\label{sec:gapequation}

Within this approximation, the Nambu self-energy is written as
\begin{align}\label{eq:McMillanSigma_app2}
\hat{\Sigma}(k)
=
\begin{pmatrix}
\Sigma_n(k) & \boldsymbol{\Delta}(k)\cdot\bsigma\, i\sigma_2 \\
-i\sigma_2\,\boldsymbol{\Delta}^*(k)\cdot\bsigma & \Sigma_n(k)
\end{pmatrix},
\end{align}
where \(\boldsymbol{\Delta}(k)\) is the spin-triplet gap function. We likewise restrict to unitary triplet states, $\boldsymbol{\Delta}^*(k)\times\boldsymbol{\Delta}(k)=0$. We assume \(\Sigma_n(-k)\approx -\Sigma_n(k)\), so that the diagonal self-energy is approximately the same in the particle and hole channels.

The corresponding Green's function is
\begin{align}
\hat G(k)
&=
\big[i\omega_n-\xi_{\vec k}\tau_3-\hat\Sigma(k)\big]^{-1}
\nonumber\\
&=
\begin{pmatrix}
G(k) & \vec F(k)\cdot\bsigma\, i\sigma_2\\
-i\sigma_2\,\vec F^*(k)\cdot\bsigma & -G(-k)
\end{pmatrix},
\label{eq:Gmatrix_app2}
\end{align}
with
\begin{subequations}\label{eq:GFdef_app2}
\begin{align}
G(k)
&=
\frac{i\omega_n-\Sigma_n(k)+\xi_{\vec k}}
{\big(i\omega_n-\Sigma_n(k)\big)^2-\xi_{\vec k}^2-|\boldsymbol{\Delta}(k)|^2}, \\
\vec F(k)
&=
\frac{\boldsymbol{\Delta}(k)}
{\big(i\omega_n-\Sigma_n(k)\big)^2-\xi_{\vec k}^2-|\boldsymbol{\Delta}(k)|^2}.
\end{align}
\end{subequations}

The gap equation follows from stationarity of the Luttinger--Ward functional with respect to \(\Delta_i^*(k)\):
\begin{equation}\label{eq:gapeq1_app2}
0
=
\frac{\delta\Omega_{\LW}}{\delta\Delta_i^*(k)}
=
\frac{\delta\Omega_{\rm el}}{\delta\Delta_i^*(k)}
+
\int_q
\Tr\!\left[
\hat\Gamma(q)\frac{\delta\hat\chi(q)}{\delta\Delta_i^*(k)}
\right].
\end{equation}
To simplify the notation, define
\begin{equation}\label{eq:Kij_app2}
K_{ij}(q)
\equiv
\big[
\Tr[\hat\Gamma(q)]-\hat\Gamma(q)-\hat\Gamma^T(q)
\big]_{ij}.
\end{equation}
Carrying out the functional derivatives then gives
\begin{subequations}\label{eq:gapeq2_app2}
\begin{align}
0
&=
2\,\Delta_i(k)\,G(k)
\Bigg[
\int_q \Tr[\hat\Gamma(q)]\,G(k+q)+\Sigma_n(k)
\Bigg]
\label{eq:line1_app2}
\\
&\quad
+
\Bigg[
\int_q\sum_{j=1}^3 K_{ij}(q)\,F_j(k+q)
-
\Delta_i(k)
\Bigg]
\label{eq:line2_app2}
\\
&\quad
+
\Delta_i(k)\sum_{j=1}^3 F_j(k)
\Bigg[
\int_q\sum_{\ell=1}^3 K_{j\ell}(q)\,F_\ell^*(k+q)
-
\Delta_j^*(k)
\Bigg]
\label{eq:line3_app2}
\\
&\quad
+
\Delta_i(k)\sum_{j=1}^3 F_j^*(k)
\Bigg[
\int_q\sum_{\ell=1}^3 K_{j\ell}(q)\,F_\ell(k+q)
-
\Delta_j(k)
\Bigg].
\label{eq:line4_app2}
\end{align}
\end{subequations}

If the first line were absent, the familiar BCS-like condition
\begin{equation}\label{eq:BCSlike_app2}
\Delta_i(k)
=
\int_q\sum_{j=1}^3 K_{ij}(q)\,F_j(k+q),
\end{equation}
would be a solution of Eq.~\eqref{eq:gapeq2_app2}, because it makes the remaining bracketed terms vanish. Both the anomalous propagator and the interaction kernel would still depend on the normal-state self-energy. The term in Eq.~\eqref{eq:line1_app2}, however, shows that even at this McMillan level there remains an additional feedback contribution associated with the superconductivity-induced change in the diagonal self-energy. Thus, once feedback is treated systematically, it enters the gap equation in a more intricate way than one would infer from a purely BCS-like renormalization of the pairing kernel.

Indeed, in the normal state,
\begin{equation}\label{eq:Sigmanormal_app2}
\Sigma_n(k)  =-\int_q \Tr[\hat\Gamma_n(q)]\,G_n(k+q),
\end{equation}
so that
\begin{align}\label{eq:Sigmachange_app2}
& \quad \Sigma_n(k)+\int_q \Tr[\hat\Gamma(q)]\,G(k+q)  \nonumber \\
&=
\int_q \Tr[\hat\Gamma(q)]\,G(k+q)
-
\int_q \Tr[\hat\Gamma_n(q)]\,G_n(k+q).
\end{align}
The combination appearing in Eq.~\eqref{eq:line1_app2} therefore directly measures the change in the diagonal self-energy caused by pairing.

Closely related feedback effects were discussed by Kuroda in the context of superfluid $^3$He, where corrections to the gap equation arising from changes in the diagonal self-energy were analyzed in order to understand the feedback of pairing on the magnetic-fluctuation contribution~\cite{Kuroda1974,Kuroda1975}.
Kuroda showed that the leading-order feedback effects derived from his gap equations are consistent with the corresponding corrections inferred from the free energy. The derivation in Refs.~\onlinecite{Kuroda1974,Kuroda1975}, however, is not especially transparent. The present formulation makes this relation more direct: both the gap equation and the corresponding free-energy correction are obtained from the same Luttinger--Ward functional. In this way one can see explicitly that the feedback terms appearing in Eq.~\eqref{eq:gapeq2_app2} are compatible with the free-energy correction discussed in the main text.

Near $T_c$, Eq.~\eqref{eq:gapeq2_app2} may be linearized in the gap, in which case only Eq.~\eqref{eq:line2_app2} survives. This yields the McMillan-type transition temperature
\begin{equation}\label{eq:mcmillan_app2}
T_c
\approx
\frac{2e^\gamma}{\pi}\,\widetilde\omega_c
\exp\!\left[-\frac{1+\lambda_Z}{\lambda_\Delta}\right],
\end{equation}
where \(\widetilde\omega_c=\omega_c/(1+\lambda_Z)\), \(\gamma\approx0.577\) is Euler's constant, and \(\lambda_\Delta\) is the pairing strength obtained by projecting the kernel \(K_{ij}(q)\) onto the dominant pairing channel in the normal state. The mass-renormalization parameter \(\lambda_Z\) is defined through
\begin{equation}
\Sigma_n(k)\approx -i\omega_n\lambda_Z,
\end{equation}
so that \((i\omega_n-\Sigma_n)\approx(1+\lambda_Z)i\omega_n\).

Although the McMillan-level approximation already goes beyond BCS theory, it remains much simpler than the full Eliashberg system because the diagonal self-energy is not updated self-consistently below \(T_c\). For determining \(T_c\), this level of approximation is often sufficient, since only the linearized gap equation is required. For the detailed structure of feedback effects below \(T_c\), however, the full Eliashberg framework is the more natural starting point. The McMillan-level treatment is nevertheless useful because it makes clear that feedback does not disappear once one relaxes full self-consistency; rather, it survives in a less compact but still systematic form.

\subsection{Derivation of the McMillan-level gap equation}
\label{app:derivations}

We now outline the derivation of Eq.~\eqref{eq:gapeq2_app2}. Starting from the stationarity condition in Eq.~\eqref{eq:gapeq1_app2}, we evaluate separately the derivatives of \(\Omega_{\rm el}\) and \(\hat\chi\) with respect to \(\Delta_i^*(k)\).

\paragraph*{1. Electronic contribution.}

We first consider \(\delta\Omega_{\rm el}/\delta\Delta_i^*(k)\). Using the matrix form of the self-energy in Eq.~\eqref{eq:McMillanSigma_app2}, one finds
\begin{equation}
\frac{\delta\Omega_{\rm el}}{\delta\Delta_i^*(k)}
=
-\frac12\int_p
\Tr\!\left[
\hat\Sigma(p)\hat G(p)
\frac{\delta\hat\Sigma(p)}{\delta\Delta_i^*(k)}
\hat G(p)
\right].
\label{eq:dOel1_app2}
\end{equation}
Since
\begin{equation}
 \frac{\delta \hat\Sigma(p)}{\delta\Delta_i^*(k)}  
=
\begin{pmatrix}
0 & 0\\
-i\sigma_2\sigma_i & 0
\end{pmatrix}
\delta_{p,k},
\end{equation}
Here $\delta_{p,k}$ denotes the delta function normalized with respect to $\int_p$, so that $\int_p\delta_{p,k}f(p)=f(k)$.
the momentum integral in Eq.~\eqref{eq:dOel1_app2} can be carried out immediately, giving
\begin{align}
& \quad \frac{\delta\Omega_{\rm el}}{\delta\Delta_i^*(k)}    \nonumber \\
&=
-\Bigg\{
\Sigma_n(k)\big[G(k)-G(-k)\big]F_i(k)
-\Delta_i(k)G(k)G(-k)
\nonumber\\
&
+\sum_{\ell,m,s=1}^3
\Delta_s^*(k)F_\ell(k)F_m(k)
\big(
\delta_{s\ell}\delta_{im}
+\delta_{sm}\delta_{\ell i}
-\delta_{m\ell}\delta_{is}
\big)
\Bigg\}.
\label{eq:dOel2_app2}
\end{align}
Using the unitarity condition stated above,
\begin{equation}
\boldsymbol{\Delta}^*(k)\times\boldsymbol{\Delta}(k)=0,
\end{equation}
the last term simplifies, and Eq.~\eqref{eq:dOel2_app2} becomes
\begin{align}
\frac{\delta\Omega_{\rm el}}{\delta\Delta_i^*(k)}
&=
-\Bigg\{
2\Sigma_n(k)G(k)F_i(k)
+\Delta_i(k)\big[-G(k)G(-k) 
\nonumber\\
& 
-\vec F^*(k)\!\cdot\!\vec F(k)\big]
+2\Delta_i(k)\,\vec F^*(k)\!\cdot\!\vec F(k)
\Bigg\}.
\end{align}
Finally, using
\begin{align}
 - [G(k)G(-k) + \vec F^*(k)\!\cdot\!\vec F(k) ]   =
\frac{1}{(i\omega_n-\Sigma_n(k))^2-\xi_{\vec k}^2-|\boldsymbol{\Delta}(k)|^2},
\end{align}
together with Eq.~\eqref{eq:GFdef_app2}, we obtain
\begin{equation}
\frac{\delta\Omega_{\rm el}}{\delta\Delta_i^*(k)}
=
-\Bigg\{
2\Sigma_n(k)G(k)F_i(k)
+F_i(k)
+2\Delta_i(k)\,\vec F^*(k)\!\cdot\!\vec F(k)
\Bigg\}.
\label{eq:dOel_final_app2}
\end{equation}

\paragraph*{2. Derivative of the susceptibility.}

We next evaluate the derivative of the susceptibility tensor. Starting from Eq.~\eqref{eq:chisc} in the main text, the required functional derivatives are
\begin{subequations}\label{eq:dGFdd_app2_revised}
\begin{align}
\frac{\delta G(p)}{\delta\Delta_i^*(k)}
&=
G(k)F_i(k)\,\delta_{p,k}, \\
\frac{\delta F_j(p)}{\delta\Delta_i^*(k)}
&=
F_j(k)F_i(k)\,\delta_{p,k}, \\
\frac{\delta F_j^*(p)}{\delta\Delta_i^*(k)}
&=
\Bigg[
\frac{\delta_{ij}}{(i\omega_n-\Sigma_n(k))^2-\xi_{\vec k}^2-|\boldsymbol{\Delta}(k)|^2}   \nonumber \\
& +F_j^*(k)F_i(k)
\Bigg]\delta_{p,k}.
\end{align}
\end{subequations}
Substituting these into Eq.~\eqref{eq:chisc}, one finds
\begin{widetext}
\begin{align}
\frac{\delta\chi_{\alpha\beta}(q)}{\delta\Delta_i^*(k)}
&=
-\delta_{\alpha\beta}F_i(k)G(k)\big[G(k+q)+G(k-q)\big]
\nonumber\\
&\quad
+\delta_{\alpha\beta}\sum_{j=1}^3
\Bigg\{
F_i(k)\big[F_j(k)F_j^*(k+q)+F_j^*(k)F_j(k-q)\big]
+\frac{\delta_{ij}}{(i\omega_n-\Sigma_n(k))^2-\xi_{\vec k}^2-|\boldsymbol{\Delta}(k)|^2}F_j(k-q)
\Bigg\}
\nonumber\\
&\quad
-F_i(k)\big[F_\alpha(k)F_\beta^*(k+q)+F_\beta^*(k)F_\alpha(k-q)\big]
-\frac{\delta_{i\beta}}{(i\omega_n-\Sigma_n(k))^2-\xi_{\vec k}^2-|\boldsymbol{\Delta}(k)|^2}F_\alpha(k-q)
\nonumber\\
&\quad
-F_i(k)\big[F_\beta(k)F_\alpha^*(k+q)+F_\alpha^*(k)F_\beta(k-q)\big]
-\frac{\delta_{i\alpha}}{(i\omega_n-\Sigma_n(k))^2-\xi_{\vec k}^2-|\boldsymbol{\Delta}(k)|^2}F_\beta(k-q).
\label{eq:dchi_app2}
\end{align}
\end{widetext}

\paragraph*{3. Fluctuation contribution and final result.}

Substituting Eq.~\eqref{eq:dchi_app2} into the second term of Eq.~\eqref{eq:gapeq1_app2}, and using that \(\hat\Gamma(q)\) is even in \(q\) for pairing states that preserve inversion and time-reversal symmetry, we obtain
\begin{widetext}
\begin{align}
\int_q
\Tr\!\left[
\hat\Gamma(q)\frac{\delta\hat\chi(q)}{\delta\Delta_i^*(k)}
\right]
&=
-2F_i(k)G(k)\int_q \Tr[\hat\Gamma(q)]\,G(k+q)
+\frac{1}{(i\omega_n-\Sigma_n(k))^2-\xi_{\vec k}^2-|\boldsymbol{\Delta}(k)|^2}
\int_q\sum_{j=1}^3 K_{ij}(q)\,F_j(k+q)
\nonumber\\
&\quad
+F_i(k)\sum_{j=1}^3F_j(k)
\int_q\sum_{\ell=1}^3 K_{j\ell}(q)\,F_\ell^*(k+q)
+F_i(k)\sum_{j=1}^3F_j^*(k)
\int_q\sum_{\ell=1}^3 K_{j\ell}(q)\,F_\ell(k+q),
\label{eq:dchi_final_app2}
\end{align}
\end{widetext}
where
\begin{equation}
K_{ij}(q)\equiv
\big[
\Tr[\hat\Gamma(q)]-\hat\Gamma(q)-\hat\Gamma^T(q)
\big]_{ij}.
\end{equation}

Combining Eq.~\eqref{eq:dchi_final_app2} with the electronic contribution in Eq.~\eqref{eq:dOel_final_app2}, and multiplying through by the common denominator
$
(i\omega_n-\Sigma_n(k))^2-\xi_{\vec k}^2-|\boldsymbol{\Delta}(k)|^2,
$
one arrives directly at Eq.~\eqref{eq:gapeq2_app2}.

 \bibliography{Feedback}
 
\end{document}